\documentclass[letterpape,11pt]{article}
\pdfoutput=1 

\usepackage{jheppub} 

\usepackage{amsmath}
\usepackage{slashed}
\usepackage{amssymb}
\usepackage{mathtools}
\usepackage{graphicx}
\usepackage{subfigure}
\usepackage{setspace}
\usepackage{sidecap}
\usepackage{color}
\usepackage{hyperref}
\usepackage{tabu} 
\usepackage{hyperref}
\usepackage[export]{adjustbox}
\usepackage{tensor}
\usepackage{enumerate}

\newcommand{\lb}{\Big{\lbrack}}
\newcommand{\rb}{\Big{\rbrack}}
\newcommand{\lp}{\Big{(}}
\newcommand{\rp}{\Big{)}}
\newcommand{\lbc}{\Big{\lbrace}}
\newcommand{\rbc}{\Big{\rbrace}}
\newcommand{\nn}{\nonumber}

\newcommand{\Bvert}{{\vert}}
\newcommand{\Rangle}{{\rangle}}
\newcommand{\Langle}{{\langle}}
\newcommand{\bmat}[1]{{\boldsymbol{#1}}}

\newcommand{\dsdqde}{\frac{d\sigma}{ de_1 de_2 d\pmb{q}_{T} } }
\newcommand{\dsdqdx}{\frac{d\sigma}{dx dQ^2 d\pmb{q}_{T}}}
\newcommand{\ecut}{e_{\text{cut}}}
\newcommand{\zcut}{z_{\text{cut}}}
\newcommand{\be}{\begin{equation}}
\newcommand{\ee}{\end{equation}}
\newcommand{\bea}{\begin{eqnarray}}
\newcommand{\eea}{\end{eqnarray}}
\renewcommand{\pmb}[1]{\boldsymbol{#1}}
\newcommand{\dijets}{\text{2\;jets}}

\newcommand{\fig}[1]{fig.~\ref{fig:#1}}

\newcommand{\cut}{\mathrm{cut}}

\title{Probing Transverse-Momentum Distributions With Groomed Jets}
\date{\today}
\author[a]{Daniel Gutierrez-Reyes,}
\author[b]{Yiannis Makris,}
\author[b]{Varun Vaidya,}
\author[a]{Ignazio Scimemi,}
\author[c,d]{and Lorenzo Zoppi\;}
\affiliation[a]{Departamento de F\'{i}sica Te\'{o}rica and IPARCOS, Universidad Complutense de Madrid (UCM), E-28040 Madrid, Spain}
\affiliation[b]{Theoretical Division, MS B283, Los Alamos National Laboratory, Los Alamos, NM 87545, USA }
\affiliation[c]{Institute for Theoretical Physics Amsterdam and Delta Institute for Theoretical Physics University of Amsterdam, Science Park 904, 1098 XH Amsterdam, The Netherlands}
\affiliation[d]{Nikhef, Theory Group, Science Park 105, 1098 XG, Amsterdam, The Netherlands}

\emailAdd{dangut01@ucm.es}
\emailAdd{yiannis@lanl.gov}
\emailAdd{vvaidya@lanl.gov}
\emailAdd{ignazios@fis.ucm.es}
\emailAdd{L.Zoppi@uva.nl}

\preprint{Nikhef: 2019-034, LA-UR-19-31102}

\abstract{We present the transverse momentum spectrum of groomed jets in di-jet events for $e^+e^-$ collisions and semi-inclusive deep inelastic scattering (SIDIS). The jets are groomed using a soft-drop grooming algorithm which helps in mitigating effects of non-global logarithms and underlying event. At the same time, by reducing the final state hadronization effects, it provides a clean access to the non-perturbative part of the  evolution of transverse momentum dependent (TMD) distributions.  In SIDIS experiments we look at the transverse momentum of the groomed jet measured w.r.t. the incoming hadron in the Breit frame. Because the final state hadronization effects are significantly reduced, the SIDIS case allows to probe the TMD parton distribution functions. We discuss  the sources of non-perturbative effects in the low transverse momentum region including novel (but small) effects that arise due to grooming. We derive a factorization theorem within SCET and resum any large logarithm in the measured transverse momentum up to NNLL accuracy using the $\zeta$-prescription as implemented in the \texttt{artemide}  package and provide a comparison with simulations.}

\begin{document}
\maketitle

\section{Introduction}
\label{sec:intro}

The understanding of the  structure of nucleons is one of the most important and interesting research subject in modern nuclear physics. The ultimate goal would be to have a complete  description of quarks/gluons position and momenta inside a hadron, which is not easy because  of the entanglement  of initial/final states in all hadronic processes. In  order to properly define all hadron constituent contributions, the cross sections  should be factorized in some region of the  phase space into properly defined hadronic matrix elements. Here we will consider the transverse momentum dependent distributions (TMD), which appear in the factorization of  several processes like Drell-Yan, semi-inclusive deep-inelastic-scattering (SIDIS) and $e^+e^-$ hadron production \cite{Collins:2011zzd,GarciaEchevarria:2011rb,Echevarria:2014rua,Gaunt:2014ska,Vladimirov:2017ksc}. Drell-Yan processes directly test the TMD parton distribution functions (TMDPDF), while in SIDIS cross sections the TMDPDFs are coupled to a TMD fragmentation function in the final state. Finally in $e^+e^-$ hadron production only the TMD fragmentation is present. Because of the factorization theorem, the TMDs have several universal features like  rapidity and renormalization scale evolution, which should be also tested including their (universal) non-perturbative part. Recently some of us have considered the possibility to define a jet-TMD, replacing a final state hadron with a jet \cite{Neill:2016vbi,Gutierrez-Reyes:2018qez,Gutierrez-Reyes:2019vbx} in SIDIS and $e^+e^-$ processes. The check of this possibility has revealed that standard jet definitions are compatible with a factorization theorem only in the case of small enough radii, which is a not obvious experimental condition in  the planned electron-hadron collider like EIC or LHeC. Instead large jet-radii need a specific definition of jet, which allows soft radiation to be independent of radius. In \cite{Gutierrez-Reyes:2018qez,Gutierrez-Reyes:2019vbx} this was achieved using the winner-take-all (WTA) axis \cite{Bertolini:2013iqa}, and the perturbative calculations were done with a precision similar to the case of fragmenting hadrons.

\begin{figure}[t!]
  \centerline{\includegraphics[width =  \textwidth]{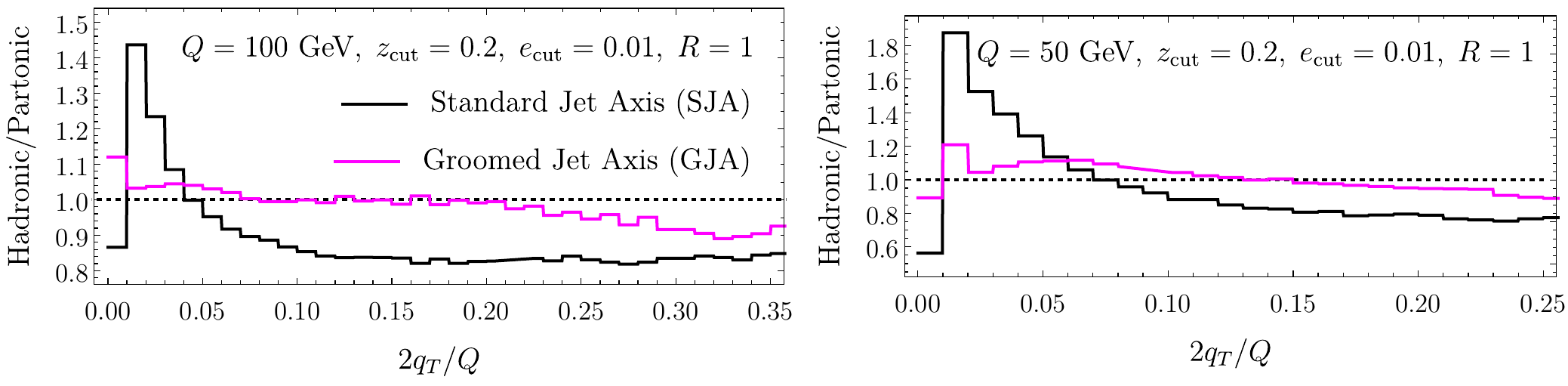}}
  \caption{Hadronization effects in a typical $e^+e^-\to \dijets$ from \textsc{Pythia} 8~\cite{Sjostrand:2006za,Sjostrand:2007gs}.
    at center of mass values, \textsc{Left}:  $Q=100$ GeV,  \textsc{Right}: $Q=50$ GeV.}
  \label{fig:hadronization}
\end{figure}

In this work we consider the possibility of groomed jets in SIDIS or $e^+e^-\to \dijets$. Developments in jet substructure have shown that applying a grooming algorithm to a jet, using for example the so called ``soft-drop'' procedure, robustly removes the contamination from both underlying event and non-global correlations. Since this process essentially removes wide angle soft radiation, retaining only a collinear core, it also dramatically reduces hadronization effects (see fig.~\ref{fig:hadronization}), thus allowing an easier access to the TMD non-perturbative physics which we want to probe. Groomed jets with an identified light/heavy hadron in the jet were also proposed as probes of TMD evolution and distribution in \cite{Makris:2017arq,Makris:2018npl}. The residual non-perturbative effects contain pieces that depend on the soft-drop grooming procedure and require careful analysis as was pointed out in \cite{Hoang:2019ceu}. In addition, with the use of soft-drop we can derive factorization theorems for large jet radius ($R\sim 1$), which we consider to be the relevant case for low energy experiments, such as EIC. In order to focus on collimated jet configurations, we also impose an upper cutoff in the groomed jet invariant mass.\footnote{Note that the small transverse momentum constraint does not necessarily ensure collimated configurations since topologies with two or more widely separated sub-jets are also permitted.} This constraint allows us to derive a factorization theorem involving the same  universal soft function that appears in traditional hadronic TMD, and which is independent of the jet radius for $R\sim 1$. This is a key feature for groomed jets and it is necessary for the universality of TMDs and for this reason, in this paper, we only consider $R\sim1$. The cutoff is imposed using groomed jet-thrust, $e \equiv (m/Q)^2$, where $m$ is the groomed invariant mass and $Q$ is the center of mass energy. This allows us to introduce a single cutoff parameter, $\ecut$, independent of the jet energy or transverse momentum.

The paper is organized as follows. In sec.~\ref{sec:main-1}, we give a review of soft-drop and discuss the factorization of the $e^+e^-\to \dijets$, transverse momentum decorrelation within SCET and give detailed comparisons of our NNLL accurate prediction with simulations for this observable. In sec.~\ref{DIS-jet}, we consider the factorization for the corresponding observable in DIS. We carefully enumerate all the non-perturbative corrections and discuss their universality in sec.~\ref{hadronization}. We conclude in sec.~\ref{conclusion}. The details of the one loop calculations, resummation, and evolution are provided in the appendix.

\section{Di-jet events in electron-positron colliders}
\label{sec:main-1}
In this section we discuss the measurement of momentum de-correlation in electron-positron colliders. We identify events with two final state jets and we consider the transverse momentum of one jet w.r.t. the other. The measurement that we are considering in this work is a generalization of the di-hadron momentum de-correlation,
\begin{equation}
  \bmat{q}_T = \frac{\bmat{p}_{T h_1}}{z_1} +  \frac{\bmat{p}_{T h_2}}{z_2} 
  \label{observable}
\end{equation}
where one or both of the identified hadrons is replaced by a jet, defined through an infrared-safe jet algorithm. Here $\bmat{p}_{T h_i}$ and $z_{i}$ are the transverse momentum and energy fraction of the hadron $i$ respectively. The factorization theorem is usually written for this normalized vector sum of the transverse momenta rather than just the sum of the transverse momenta. It can be verified by momentum conservation and simple geometry that the quantity $\bmat{p}_{T h_1}/z_1$ represents the transverse momentum of the radiation recoiling against the hadron w.r.t the axis defined by the hadron itself. This makes it convenient to write a factorization theorem which matches onto the standard hadron fragmentation function as explained in \cite{Makris:2017arq}.

\begin{figure}[t!]
  \centerline{\includegraphics[width =  \textwidth]{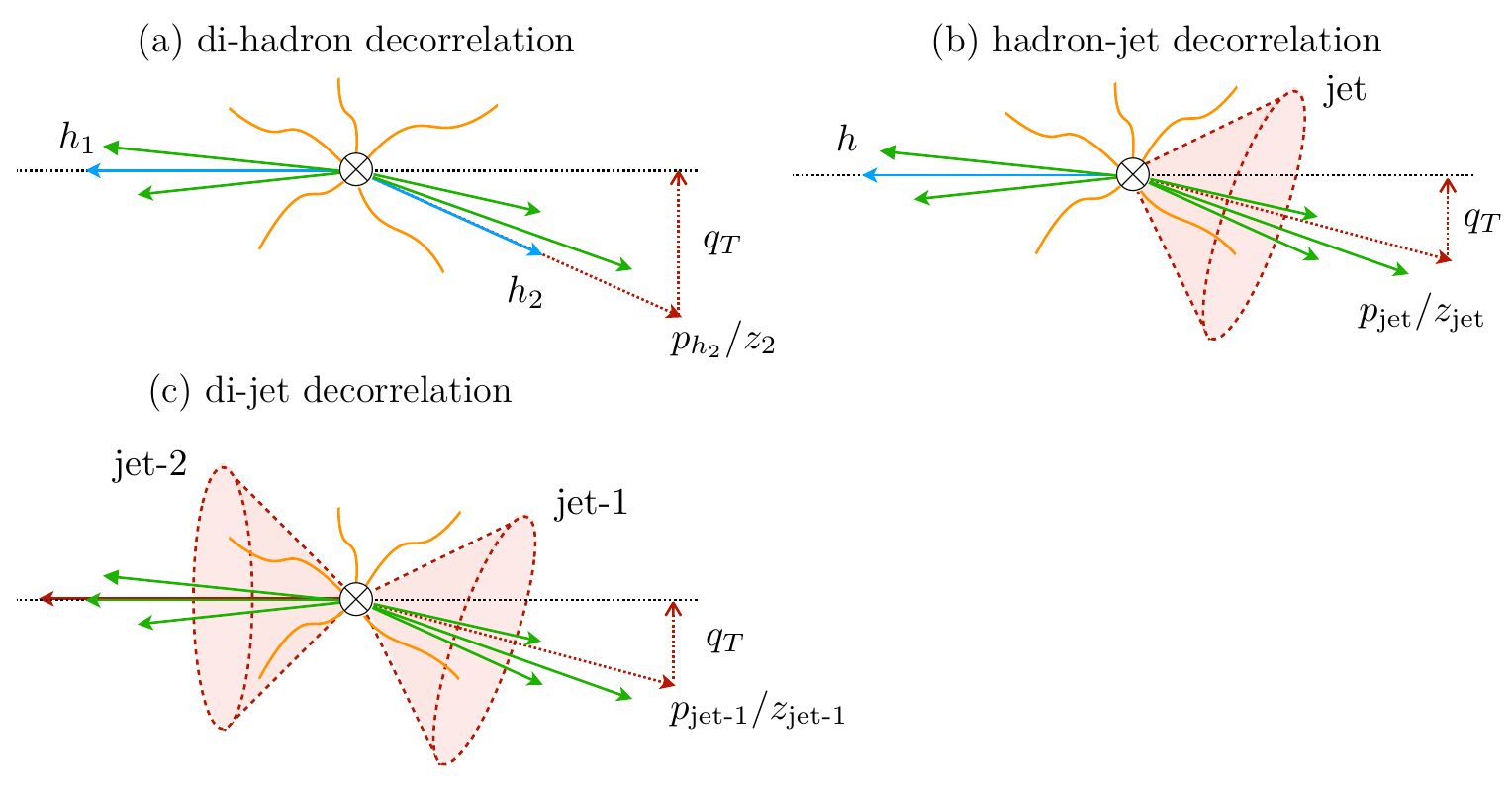}}
  \caption{Three possible transverse momentum de-correlation measurements in $e^+ e^-$ annihilation: (a) Identify two hadrons $h_1$ and $h_2$ with momenta $p_{h_1}, p_{h_2}$ and energy fractions $z_{h_1},z_{h_2}$ respectively, (b) Identify a jet and a hadron with momenta $p_{\text{jet}}$, $p_{h}$ with energy fractions $z_{\text{jet}}, z_h$, (c) Identify two jets with momenta $p_{\text{jet}_1},p_{\text{jet}_2}$ and energy fractions $z_{\text{jet}_1}, z_{\text{jet}_2}$.}
  \label{fig:measurment}
\end{figure}

We consider three possible scenarios as illustrated in fig.~\ref{fig:measurment} and we refer to them as di-hadron, hadron-jet, and di-jet momentum de-correlation.  To simplify the discussion we focus on the case  of di-jets (fig.~\ref{fig:measurment}c) and we briefly comment how our results are generalized for the case of hadron-jet de-correlation. For the case of groomed jets the observable $\bmat{q}_T$ is defined with the groomed quantities, i.e., $p_{\text{jet}}^{\mu}$ is the \emph{groomed} jet four-momentum and $z_{\text{jet}} = 2 p^{0}_{\text{jet}}/Q$. The transverse component $\bmat{p}_{T\;\text{jet}}$ is measured with respect to an axis close to the full or groomed jet axes. The exact choice of the axis only differs by power corrections. For concreteness in the results that follow we make the choice of the axis to lie along one of the groomed jets. 

Since we want to probe the non-perturbative physics, we wish to work in the small transverse momentum regime ($q_T \ll \sqrt{s}$ where $q_T \equiv \vert \pmb{q}_T \vert$). There are various ways one can define the jet axis and the choice of definition will impact the form of factorization. It was discussed in ref.~\cite{Gutierrez-Reyes:2018qez} that the standard jet axis choice suffers from factorization breakdown for large jet radius. This breakdown is due to energetic emissions at relatively wide angles. Such configurations will contribute to the small transverse momentum region when the energetic subjets are clustered in a single large radius jet. To avoid this problem in refs.~\cite{Gutierrez-Reyes:2018qez, Gutierrez-Reyes:2019vbx} the winner-take-all (WTA) axis was used instead. This way  ensures that wide angle energetic emissions induce large transverse momentum ($q_T \sim \sqrt{s}$) pushing the $q_T$ measurement away from the observable region. 

In this paper we propose, alternatively, the use of groomed jet-substructure to isolate the collimated configurations and choose the jet axis to be the groomed jet axis which is insensitive to jet boundary effects. Particularly we consider the normalized jet mass as the relevant jet-substructure observable,
\begin{equation}
  e\equiv \lp\frac{m_J}{Q}\rp^2\ .
  \label{jetMass}
\end{equation}
We shall see that imposing this constraint still allows us to capture a majority of events and hence does not significantly impact the cross-section. 
\subsection{Soft-drop: a brief review}
\label{sec:review}

The grooming procedure that we use is the soft-drop algorithm. We give here a brief review of the soft-drop groomer and eventually discuss the various hierarchies, the relevant modes and the factorization of the cross section in the next sections. 

Soft-drop grooming \cite{Larkoski:2014wba} removes contaminating soft radiation from the jet by constructing an angular ordered tree of the jet, and removing the branches at the widest angles which fail an energy requirement. The angular ordering of the jet is constructed through the Cambridge/Aachen (C/A) clustering algorithm \cite{Ellis:1993tq,Catani:1993hr,Dokshitzer:1997in,Wobisch:1998wt,Wobisch:2000dk}. As soon as a branch is found that passes the test, it is declared the groomed jet, and all the constituents of the branch are the groomed constituents. At the end of the grooming procedure only the narrow energetic core remains from the original jet. Since at large angles all collinear energetic radiation is to be found at the center of the jet, no cone is actually imposed to enclose this core. One simply finds the branch whose daughters are sufficiently energetic. Formally the daughters could have any opening angle, though their most likely configuration is collinear.

The strict definition of the algorithm is as follows. Given an ungroomed jet (which itself is identified first using a suitable algorithm such as the anti-$k_T$, \cite{Cacciari:2008gp}), first we build the clustering history by starting with a list of particles in the jet. At each stage we merge the two particles within the list that are closest in angle\footnote{This merging is usually taken to be summing the momenta of the particles, though one could use winner-take-all schemes \cite{Salam:WTAUnpublished, Bertolini:2013iqa,Larkoski:2014uqa}.}. This gives a pseudo-particle, and we remove the two daughters from the current list of particles, replacing them with the merged pseudo-particle. This is repeated until all particles are merged into a single parent. Then we open the tree back up working backwards so that at each stage of the declustering, we have two branches available, label them $i$ and $j$. We require:
\begin{align}
  \label{SD:condition}
  \frac{\text{min}\{E_i,E_j\}}{E_i+E_j}> \zcut \lp \frac{\theta_{ij}}{R} \rp^{\beta},
\end{align}
where $\zcut$ is the modified mass drop parameter, $\beta$ is the parameter which controls the angularities, $\theta_{ij}$ is the angle between $i^{th}$ and $j^{th}$ particle, $R$ is the jet radius and $E_i$ is the energy of the branch $i$. If the two branches fail this requirement, the softer branch is removed from the jet, and we decluster the harder branch, once again testing eq.~(\ref{SD:condition}) within the hard branch. The pruning continues until we have a branch that when declustered passes the condition eq.~(\ref{SD:condition}). All particles contained within this branch whose daughters are sufficiently energetic constitute the groomed jet. Intuitively we have identified the first genuine collinear splitting.

For a hadron-hadron collision, one uses the transverse momentum $(p_T)$ with respect to the beam for the condition of eq.~(\ref{SD:condition}),
\begin{align}\label{SD:condition_pp}
  \frac{\text{min}\{p_{Ti},p_{Tj}\}}{p_{Ti}+p_{Tj}}> \zcut \lp \frac{\theta_{ij}}{R} \rp ^{\beta}.
\end{align}

We formally adopt the power counting $\zcut \ll 1$, though typically one chooses $\zcut \sim 0.1$. See \cite{Marzani:2017mva} for a study on the magnitude of the power corrections with respect to $\zcut$ for jet mass distributions. To be specific, in this paper we consider only the case $\beta = 0$.  Note that for $\beta =0$ the energy of the groomed jet constituents is a collinear unsafe observable~\cite{Larkoski:2014wba, Larkoski:2015lea}, however, the additional constraint of the measured transverse momentum $\bmat{q}_T$ provides a physical collinear cutoff in a similar way a jet shape measurement does. For detailed discussion on this we refer to appendix~\ref{app:IRCsafe}.

\subsection{Hierarchies, modes, and factorization}
\label{sec:hierarchy}

In order to compute the transverse momentum de-correlation $\bmat{q}_T$, defined in eq.~(\ref{observable}), for two groomed jets in di-jet events in $e^+ e^-$ annihilation (fig.~\ref{fig:measurment} (c))
we  are going to impose a normalized jet mass measurement as defined in 
eq.~(\ref{jetMass}) on both jets. The other parameters that enter our cross section are the soft-drop parameters $\zcut \sim 0.1$,  $\beta=0$. Ultimately we are going to integrate over the jet mass measurement up to an appropriate (but still small) cut-off value $\ecut$. \\

We have a rich spectrum of possible hierarchies of momenta,  which are all consistent with maintaining $ q_T/Q , \ecut$, $\zcut\ll 1 $. We have that   $ q_T/Q , \ecut$, $\zcut $ are now expansion parameters in the effective field theory (EFT), and they should be taken into account in the factorization of the process. We first list and briefly discuss these hierarchies and the corresponding factorization theorems within an EFT. The general modes that we will consider will fall into three classes. Modes that explicitly pass soft drop (usually the highly energetic collinear modes),  modes that explicitly fail soft-drop (the global soft function modes) and finally those which can live on the border and need to be tested, as to whether they pass or fail. Only the modes that pass soft-drop will contribute to $e$, while $\bmat{q}_T$ receives contributions from all radiation that fails soft-drop.

The first regime in which we are interested is 
$Q\gg Q \zcut \gg q_T \gtrsim Q\sqrt{e} \gg Q\sqrt{e \zcut}$. Here we have low values of $q_T$ which are of the order of $Q\sqrt{e}$. We identify the following modes to be relevant to the cross section:
\begin{align}
  \text{soft:}& \;\;\; p_{s}^{\mu} \sim q_T(1,1,1); \nn\\
  \text{collinear:}& \;\;\; p_{c}^{\mu} \sim Q(\lambda_c^{2},1,\lambda_c) ,\; \lambda_c = \sqrt{e},
\end{align}
and the factorization of the cross section in this region is schematically
\begin{equation}
  \label{eq:fact-II}
  \dsdqde =  H^{ij}_2(Q;\mu) \times S(\pmb{q}_T)  \otimes \mathcal{J}^{\perp}_{i}(e_1, Q,\zcut,\pmb{q}_T) \otimes \mathcal{J}^{\perp}_{j}(e_2, Q,\zcut,\pmb{q}_T).
\end{equation}
Apart from the hard factor $H$ all the other terms in this equation are affected by rapidity divergences. 
The global soft function $S$ that appears in the factorization theorem in eq.~(\ref{eq:fact-II}) (and later in the SIDIS case eq.~(\ref{eq:fact_ep})) is the universal function that is also present in the factorization theorem of
Drell-Yan, di-hadron production in electron-positron annihilation, and semi-inclusive DIS with TMDs. The operator definition of the soft function (see refs.~\cite{Collins:2011zzd,GarciaEchevarria:2011rb,Chiu:2012ir}) is given by
\begin{equation}
  \label{eq:soft}
  S(\pmb{q}_T) = \frac{1}{N_R} \text{tr} \;\Langle [ S_n^{\dag} S_{\bar{n}}](0) \delta^{(2)}(\pmb{q}_T - \pmb{\mathcal{P}}_{\perp}) [ S_{\bar{n}}^{\dag} S_{n}](0)  \Rangle\;,
\end{equation}
where $N_R = N_c$ for $S_{n/\bar{n}}$ in the fundamental and $N_c^2 -1$ for the adjoint representation of $SU(N_c)$.
This function has been calculated at NNLO in \cite{Echevarria:2015byo}. This function is responsible for the TMD evolution which is actually known up to third order
\cite{Li:2016ctv,Vladimirov:2016dll}. The power corrections to the evolution have been studied in \cite{Scimemi:2016ffw}. Because  of the universality of this soft function the non-perturbative corrections that it generates in the TMD-evolution factor are process independent \cite{Collins:2011zzd,GarciaEchevarria:2011rb,Scimemi:2016ffw}.

The soft factor provides finally a  rapidity renormalization factor for the jets which is totally analogous to the TMD case, see ref.~\cite{Echevarria:2016scs},
so that in this sense we can re-write eq.~(\ref{eq:fact-II}) as
\begin{equation}
  \label{eq:fact-IIa}
  \dsdqde =  H^{ij}_2(Q;\mu)\times \mathcal{J}^{\perp}_{i}(e_1, Q,\zcut,\pmb{q}_T;\mu,\zeta_A) \otimes \mathcal{J}^{\perp}_{j}(e_2, Q,\zcut,\pmb{q}_T;\mu,\zeta_B)\;,
\end{equation}
with $\zeta_A\zeta_B=Q^4z_\cut^4$  , which recalls clearly the all-order factorization  for the di-hadron fragmentation case  using TMD. The hadronization corrections to eq.~(\ref{eq:fact-II}-\ref{eq:fact-IIa}) are discussed in more detail in sec.~\ref{hadronization}.

\begin{figure}[t!]
  \centerline{\includegraphics[width =  \textwidth]{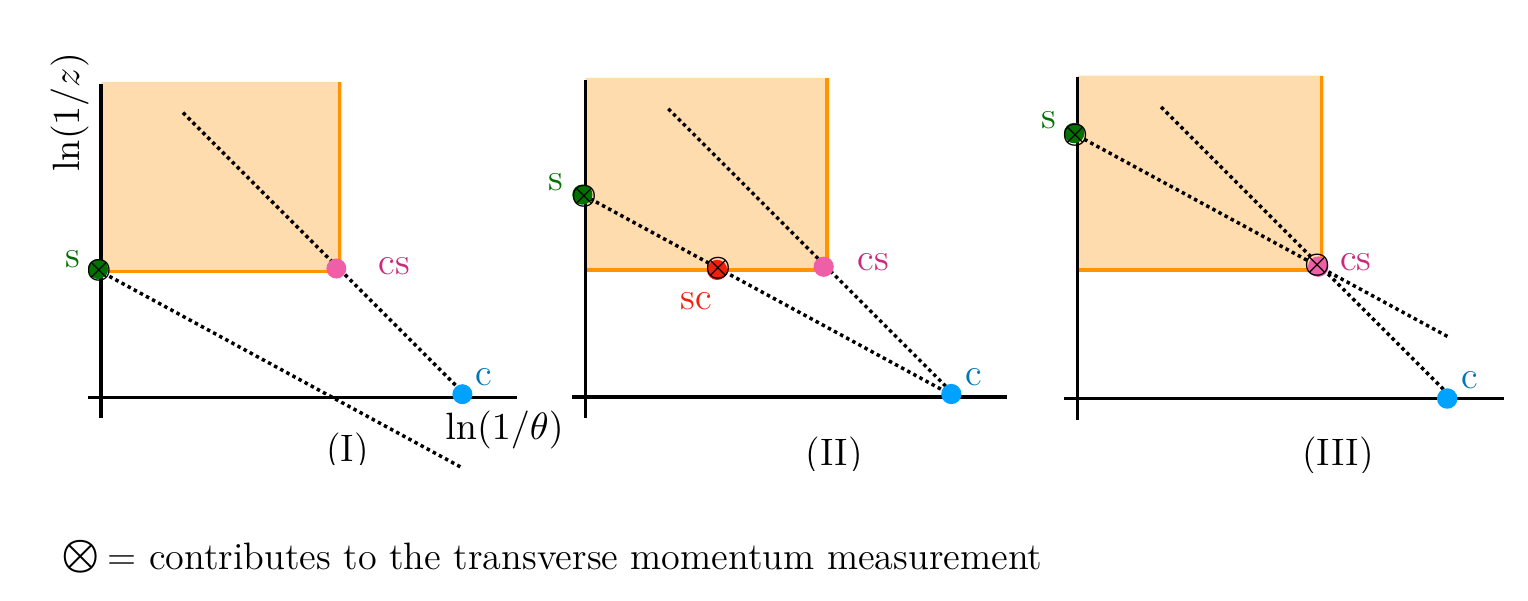}}
  \caption{Three possible hierarchies for $q_T$. Shaded region is one that fails Soft-Drop. (I) Largest $q_T \sim Qz_{\text{cut}}$. The cross section is factorized into 3 function s, cs and c. (II) The soft function s splits into two s and sc.(III)  The sc function merges with the cs function.}
  \label{fig:regions}
\end{figure}

The jet-TMD of eq.~(\ref{eq:fact-IIa})  can be re-factorized  depending on the relative magnitudes of the effective scales which  define it  so that one can identify the more complete set of modes
\begin{align}
  \text{soft:}& \;\;\; p_{s}^{\mu} \sim q_T(1,1,1); \nn\\
  \text{collinear:}& \;\;\; p_{c}^{\mu} \sim Q(\lambda_c^{2},1,\lambda_c) ,\; \lambda_c = \sqrt{e};\nn\\
  \text{soft-collinear:}& \;\;\; p_{sc}^{\mu} \sim Q\zcut(\lambda_{sc}^{2},1,\lambda_{sc}),\; \lambda_{sc} =   q_T /(Q\zcut); \nn\\
  \text{collinear-soft:}& \;\;\; p_{cs}^{\mu} \sim Q\zcut(\lambda_{cs}^{2},1,\lambda_{cs}),\; \lambda_{cs} = \sqrt{e / \zcut}
\end{align}
and we illustrate this in fig.~\ref{fig:regions}.
We start considering the limit $q_T \gtrsim Q\sqrt{e} \gg Q\sqrt{e \zcut}$, which corresponds to region II in  fig.~\ref{fig:regions},
when
the unintegrated and unsubtracted jet function, $\mathcal{J}^{\perp}_{i}$, in eq.~(\ref{eq:fact-II}) can be re-factorized into three terms,
\begin{equation}
  \label{eq:jet-fact-II}
  \mathcal{J}^{\perp}_{i}(e, Q,\zcut,\pmb{q}_T) = S_{sc,i}^{\perp}(Q\zcut,\pmb{q}_{T}) \times \int de'\; S_{cs,i}(e-e',Q\zcut) J_i(e',Q)
\end{equation}
where 
all the rapidity divergent part and transverse  momentum dependence is contained in the calculable $S_{sc,i}^{\perp}$. The subtracted and  unsubtracted jet-TMD are related by
\begin{align} \label{eq:jj}
  \mathcal{J}^{\perp}_{i}(e, Q,\zcut,\pmb{b},\mu,\zeta)=\sqrt{S(\pmb{b})} \mathcal{J}^{\perp}_{i}(e, Q,\zcut,\pmb{b})
\end{align}
where  we have expressed all the subtraction  in $\pmb{b}$-space. \footnote{Throughout the paper we will interchange between $\pmb{q}_T, \pmb{b}$ spaces for the transverse spectrum and between $e, s$ spaces for the jet mass. We use the same symbol for any function in either space. The variable we are working in should be clear from the argument of the function.}
For smaller values of $q_T$: $Q\gg Q \zcut \gtrsim Q\sqrt{e} \gg q_{T} \sim Q\sqrt{e \zcut}$, the collinear-soft and soft-collinear merge into the same mode which we still refer to as collinear-soft. The soft and collinear modes remain unchanged in their scaling compared to region II. The form of factorization theorem in eq.~(\ref{eq:fact-II}) does not change but now the corresponding jet TMDs are re-factorized as (see region III in  fig.~\ref{fig:regions}),
\begin{equation}
  \label{eq:jet-fact-III}
  \mathcal{J}^{\perp}_{i}(e, Q,\zcut,\pmb{q}_{T}) = \int de'\; S_{cs,i}^{\perp}(e-e',Q\zcut,\pmb{q}_T) J_i(e',Q).
\end{equation}
Several of the parameters in the differential cross-secion in eq.~(\ref{eq:fact-IIa}) are in practice integrated in experiments, so that it is convenient to explicitly write the cumulant (or partially integrated) distribution
\begin{equation}
  \label{eq:int-ds}
  \frac{d\sigma}{d\pmb{q}_T} (\ecut)= \int_0^{\ecut} de_1 de_2 \dsdqde \; .
\end{equation}
For  this cross section we work with the integrated jet function which depends on $\ecut$ rather than $e$,
\begin{equation}
  \mathcal{J}^{\perp}_{j}(\ecut, Q,\zcut, \bmat{q}_T;\mu,\zeta) = \int_0^{\ecut} de \; \mathcal{J}^{\perp}_{j}(e, Q,\zcut, \bmat{q}_T;\mu,\zeta)\;.
\end{equation}
and the factorization theorem for electron-positron annihilation is 
\begin{equation}
  \label{eq:fact_ee}
  \frac{d\sigma}{d\pmb{q}_T} (\ecut)= H^{ij}_2(Q;\mu)    \int \frac{d \pmb{b}}{4\pi} e^{i \pmb {b} \cdot \pmb{q}_T} \mathcal{J}^{\perp}_{i}(\ecut, Q,\zcut,\pmb b;\mu,\zeta)  \mathcal{J}^{\perp}_{j}(\ecut, Q,\zcut,\pmb b;\mu,\zeta)\;.
\end{equation}
The resummation of logarithms inside the jet-TMD implied by eq.~(\ref{eq:jet-fact-II}-\ref{eq:jet-fact-III}) is taken into account defining
the cumulant jet function as
\begin{align}
  \label{eq:cum-jet0}
  \mathcal{J}^{\perp}_{i}(\ecut, Q,\zcut,\pmb{b};\mu,\zeta)&=\sqrt{S(\pmb{b})} \mathcal{J}^{\perp}_{i}(\ecut, Q,\zcut,\pmb{b})\ ,
  \\
  \mathcal{J}^{\perp}_{i}(\ecut, Q,\zcut,\pmb{b}) &= S_{sc,i}^{\perp}(Q\zcut,\pmb{b})   \mathcal{J}_{i}(\ecut, Q,\zcut;\mu) ,
  \\
  \label{eq:cum-jet}
  \mathcal{J}_{i}(\ecut, Q,\zcut;\mu) &= \int_0^{\ecut} de \int de'\; S_{cs,i}(e-e',Q,\zcut;\mu) J_i(e',Q;\mu) 
\end{align}
and  we recall that the rapidity divergences are present  only in $S$ and $S_{sc,i}^{\perp}$, canceling in their product in eq.~(\ref{eq:cum-jet0}). With the exception of the soft-collinear function, $S_{sc}^{\perp}$, all other ingredients of the factorization are already known at least up to NLO accuracy. In app.~\ref{oneloop} we report the defining matrix elements of  each function,  we summarize  the NLO  results and we perform the NLO calculation of $S_{sc}^{\perp}$. We have performed the calculation using rapidity regulator.  The connection between rapidity regulator  and $\zeta$-parameter is outlined in app.~\ref{sec:nu-zeta}.

\begin{figure}[t!]
  \centerline{\includegraphics[width = \textwidth]{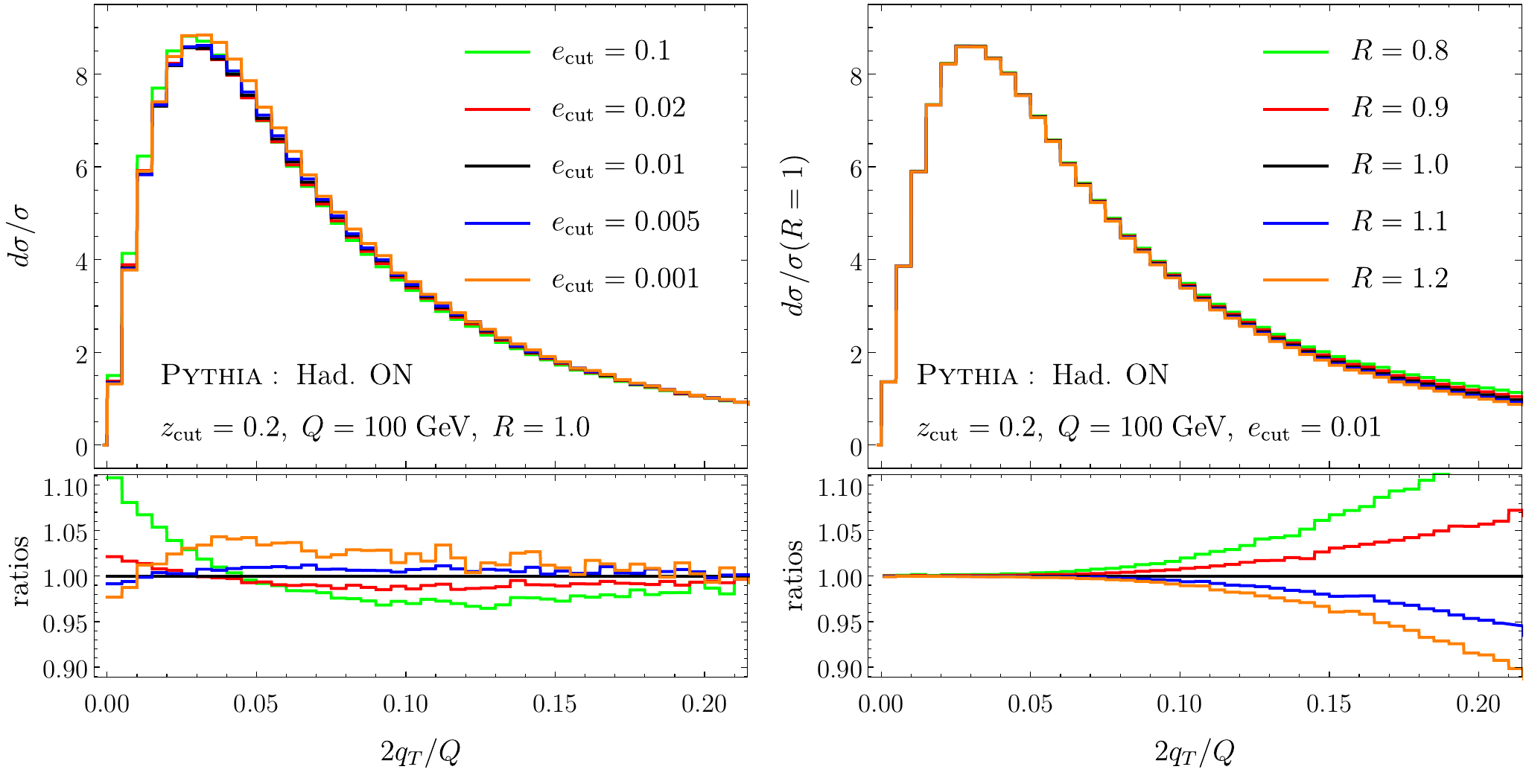}}
  \caption{\textsc{Left}: The normalized cross sections for different values of the jet mass cutoff parameter $\ecut$. We also include the corresponding ratios with respect to the case $\ecut =0.01$. \textsc{Right}: The \emph{relatively} normalized cross section for fixed $\ecut =0.01$ and for different value of the jet radius $R$. The corresponding ratios are with respect to $R=1$. }
  \label{fig:var_ecut}
\end{figure}

Finally we observe  that using monte-carlo simulations (particularly \textsc{Pythia} 8~\cite{Sjostrand:2006za,Sjostrand:2007gs}) most of the events  fall in the  kinematic regime
\begin{equation} \label{eq:hierarchy}
  Q\gg Q \zcut \gg  q_T \sim Q \sqrt{\ecut} \;. 
\end{equation}
An important consequence of the jet function refactorization in~eq.~(\ref{eq:jet-fact-II}) is that the transverse momentum dependent elements decouple from the jet mass elements. This suggests that, as long as we remain within the hierarchy of eq.~(\ref{eq:hierarchy}), then the exact mass cutoff on the invariant mass will only influence the overall normalization and not the shape of the TMD distribution. We test this observation against the monte-carlo simulations by comparing the normalized TMD distributions for various values of $\ecut$. We show the results in fig.~\ref{fig:var_ecut} (left). The jet algorithm is implemented through \textsc{FastJet}-3~\cite{Cacciari:2011ma}.
In addition we note that as long as we measure $q_T \ll Q\zcut$ and for $R \sim 1$ the shape and normalization of the cross section is independent of the choice of $R$. We also demonstrate this with the help of simulations. We simulate events at $Q = 50$ GeV and we analyze them for different values of $R\gtrsim 1$. We show the resulting distributions in fig.~\ref{fig:var_ecut} (right). Note that for that plot we preserve the relative normalizations of the curves.

\subsection{Renormalization group evolution}

The two main quantities involved in the factorization procedure carried out in previous section are the subtracted jet-TMD for which we have
\begin{align}
  \label{eq:dsevTMDs}
  &\mu\frac{d}{d\mu} \mathcal{J}^{\perp}(e,Q,z_{\rm cut},\boldsymbol b, \mu, \zeta)=\gamma_F^q(\mu,\zeta) \mathcal{J}^{\perp}(e,Q,z_{\rm cut},\boldsymbol b, \mu, \zeta),\\
  \label{eq:dsevTMDs-z}
  &\zeta \frac{d}{d\zeta} \mathcal{J}^{\perp}(e,Q,z_{\rm cut},\boldsymbol b, \mu, \zeta)=-\mathcal{D}^q(\mu,\boldsymbol b) \mathcal{J}^{\perp}(e,Q,z_{\rm cut},\boldsymbol b, \mu, \zeta),
\end{align}
where on the r.h.s. we have  considered just quark initiated jets and we have Fourier transformed  with respect to $\bmat{q}_T$ the jet functions appearing in eq.~(\ref{eq:fact-IIa}). Of course this result recalls literally the standard TMD case.

However, because of the re-factorization of $ \mathcal{J}^{\perp}$  (see eq.~(\ref{eq:jet-fact-II}-\ref{eq:jet-fact-III})) this resummation is not complete and large logarithms can still spoil  the convergence of the perturbative series. Defining $s$ as the variable conjugate to $e$ in Laplace space (see app.~\ref{transforms}) and 
\begin{align}
  G\in\left\{S_{sc}^{sub}(Qz_{\rm cut},\pmb{b}),\, S_{cs}(s,Q z_{cut}),\, J(s,Q)\right\};\quad S_{sc}^{sub}(Qz_{\rm cut},\pmb{b})=\sqrt{S(\pmb b)} S_{sc}(Qz_{\rm cut},\pmb{b})\;,
\end{align}
we have
\begin{equation}
  \label{eq:unmeasRG}
  \mu \frac{d}{d \mu}G= \gamma^{G} (\mu, \alpha_s) G= \left( \Gamma^{G} [\alpha_S] \textbf{l}_{m^2_{G}}  + \Delta \gamma^{G}[\alpha_S]\right) G,
\end{equation}
which are formally similar to the TMD case and the values of $m_G$ are reported in the appendix in tab.~\ref{tb:evolution}. The only function in $G$ which has a rapidity evolution equation is $S_{sc}^{sub}$  and it scales  like  $ \mathcal{J}^{\perp}$ in eq.~(\ref{eq:dsevTMDs-z}).
The  cusp part of eq.~(\ref{eq:unmeasRG})  is proportional to the standard cusp anomalous dimension
\begin{equation}
  \label{eq:G}
  \Gamma^{G}_{\mu}[\alpha_s] =  \frac{\Gamma^G_0}{\Gamma^{\text{cusp}}_0} \Gamma^{\text{cusp}} = \frac{\Gamma^G_0}{\Gamma^{\text{cusp}}_0}  \sum_{n=0}^{\infty} \left(\frac{\alpha_s}{4 \pi} \right)^{1+n} \Gamma^{\text{cusp}}_n,
\end{equation}
For the non-cusp part we have also a  perturbative  expansion
\begin{equation}
  \label{eq:g}
  \Delta \gamma^{G}[\alpha_S] =  \sum_{n=0}^{\infty} \left(\frac{\alpha_s}{4 \pi} \right)^{1+n} \gamma^{G}_n.
\end{equation}
The anomalous dimensions that enter in the calculations for each case are given in  app.~\ref{app:evolution}. The evolution in rapidity and factorization scales of all  quantities  can be implemented using the the $\zeta$-prescription whose general framework can be found in  ref.~\cite{Scimemi:2018xaf}.  We provide some details for the present case in the appendix.

The  resummation of   potentially large logarithms inside the jet-TMD is done  performing the evolution in Laplace space and then integrating such that we  get the cumulant before we take the inverse transform. 
In this way we resum logarithms which are associated to  $\ecut$.  All this works as follows.
Starting from eq.~(\ref{eq:cum-jet}), then taking the Laplace and consecutively the inverse transform with respect to $e$ we find
\begin{equation}
  \mathcal{J}_{i}(\ecut, Q,\zcut;\mu) = \frac{1}{2\pi i} \int_{\gamma-i \infty}^{\gamma+i\infty} ds \frac{\exp(s \ecut) -1}{s} S_{cs,i}(s,Q,\zcut;\mu) J_i(s,Q;\mu)\ .
\end{equation}
Then solving the RGE equations for the collinear-soft and jet function as described in app.~\ref{sec:RGEs}, and performing the last remaining integral over the Laplace conjugate variable $s$ we get
\begin{multline}
  \mathcal{J}_{i}(\ecut, Q,\zcut;\mu)  = \exp \lp K_{cs}(\mu,\mu_{cs})+K_{J}(\mu,\mu_{J}) \rp   S_{cs,i} (L_{cs} \to \partial_{\omega_{cs}};\mu_{cs}) J_i  (L_{J} \to \partial_{\omega_{J}};\mu_J) \\
  \lp \frac{\mu_{cs}}{ Q \sqrt{\zcut \ecut}} \rp^{2\omega_{cs}(\mu,\mu_{cs})}  \lp \frac{\mu_{J}}{ Q \sqrt{\ecut}} \rp^{2\omega_{J}(\mu,\mu_{J})} \frac{\exp(\gamma_E(\omega_{cs}(\mu,\mu_{cs})+\omega_{J}(\mu,\mu_{J})))}{\Gamma(1-\omega_{cs}(\mu,\mu_{cs})-\omega_{J}(\mu,\mu_{J}))}   \ .
\end{multline}
This is our final result for the resummed cumulant jet function. The order of logarithmic accuracy is then determined by the order of which the kernels $K_F$, $\omega_F$, and the fixed order collinear-soft and jet functions are evaluated. At this stage of the calculation the canonical scales, $\mu_{cs}$ and $\mu_J$, are not yet fixed.  This allows us to choose the scales such that potentially large logarithms are minimized in  momentum space. From the above is clear that the canonical choice of scales such as the fixed order logarithms are minimized are,
\begin{align}
  \mu_{cs} &= Q \sqrt{\zcut \ecut}\;, & \mu_J &= Q \sqrt{\ecut} \;.
\end{align}
In numerical applications one needs to perform variations around these scales in order to obtain an estimate of the theoretical uncertainty.

\subsection{Numerical results for $e^+e^-$} 
\label{sec:numerics-ee}

In this section, we provide the results of our calculation for $e^+e^-\to \dijets$  computed up to NNLL accuracy.  The implementation necessarily  needs  a choice for the rapidity scales and  we have done it using the $\zeta$-prescription as described in ref.~\cite{Scimemi:2018xaf} and adapting the  code \texttt{artemide}   to the present case. 
This consisted of performing the evolution of the transverse momentum dependent components within the \texttt{artemide}  framework, while 
for all other scales not involved  in the rapidity evolution,
i.e., the hard and jet functions,   see  app.~\ref{sec:RGEs}.

There are some important modifications to the $\zeta$-prescription framework    for our case which affect the numerics. 
One of this is that now 
$\zeta_A \zeta_B \sim Q^4 \zcut^4$ compared to the di-hadron decorrelation case where $ \zeta_A \zeta_B \sim Q^4  $.  
This  means that the effective hard scale to which the distributions are sensitive is lower. Because the TMD factorization is  valid when $q_T$ is much lower than the hard scale of the process, 
one needs that the product  $Q z_\cut$ be sufficiently high.  In our plots we have  considered the case $q_T\lesssim Q z_\cut$. Then the evolution of the jet-TMD given in eq.~(\ref{eq:unmeasRG}) is also slightly different from the standard hadron TMD, although the changes are implemented easily in  the \texttt{artemide}  code. A  one-loop check of all  anomalous dimensions is provided in app.~\ref{oneloop}.

\begin{figure}[h!]
  \centerline{\includegraphics[width =  \textwidth]{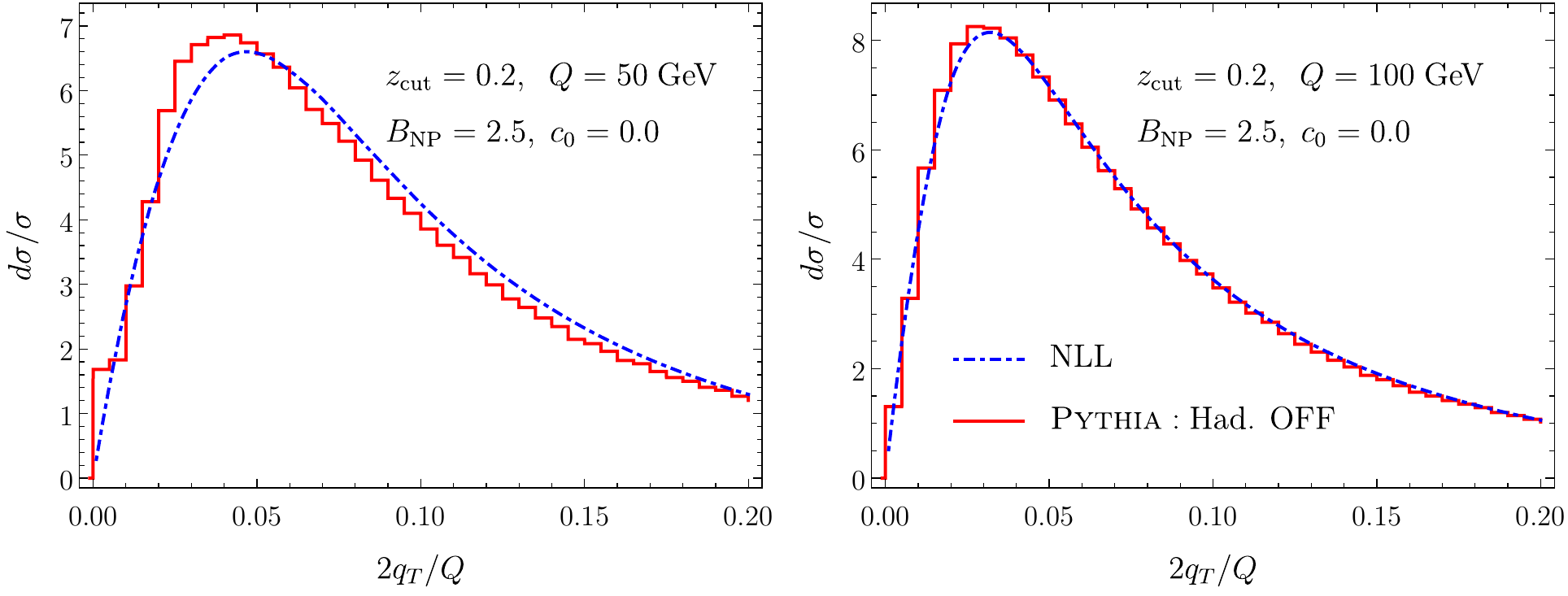}}
  \caption{Comparison of the NLL result against the partonic shower of \textsc{Pythia} 8 for $R=1$ and $\ecut =0.01$ for two different center of mass energies, \textsc{Left}: 50 GeV, \textsc{Right}: 100 GeV.}
  \label{fig:e+e-MC}
\end{figure}
In fig.~\ref{fig:e+e-MC} we compare our analytic result for NLL cross section (normalized) against \textsc{Pythia} simulations for $Q=50$ and 100 GeV.  For the purposes of comparison we turn hadronization off in the simulation and we compare against our purely perturbative result. The perturbative calculation depends on the parameter $B_{\text{NP}}$ which in practice implements a cutoff in the inverse Laplace transform such that the soft scale, that behaves as $1/b$, does not hit the Landau pole. As long as we choose this parameter such that convergence of the integral is reached before the cutoff, then the perturbative result is not much sensitive to the value of $B_{\text{NP}}$. Although, as we now discuss, the theoretical uncertainty of the cross section for these energies at NLL is quite large, we find very good agreement with the simulations for the canonical choice of scales (i.e., central line in fig.~\ref{fig:e+e-MC}).

\begin{figure}[t!]
  \centerline{\includegraphics[width =  0.517\textwidth]{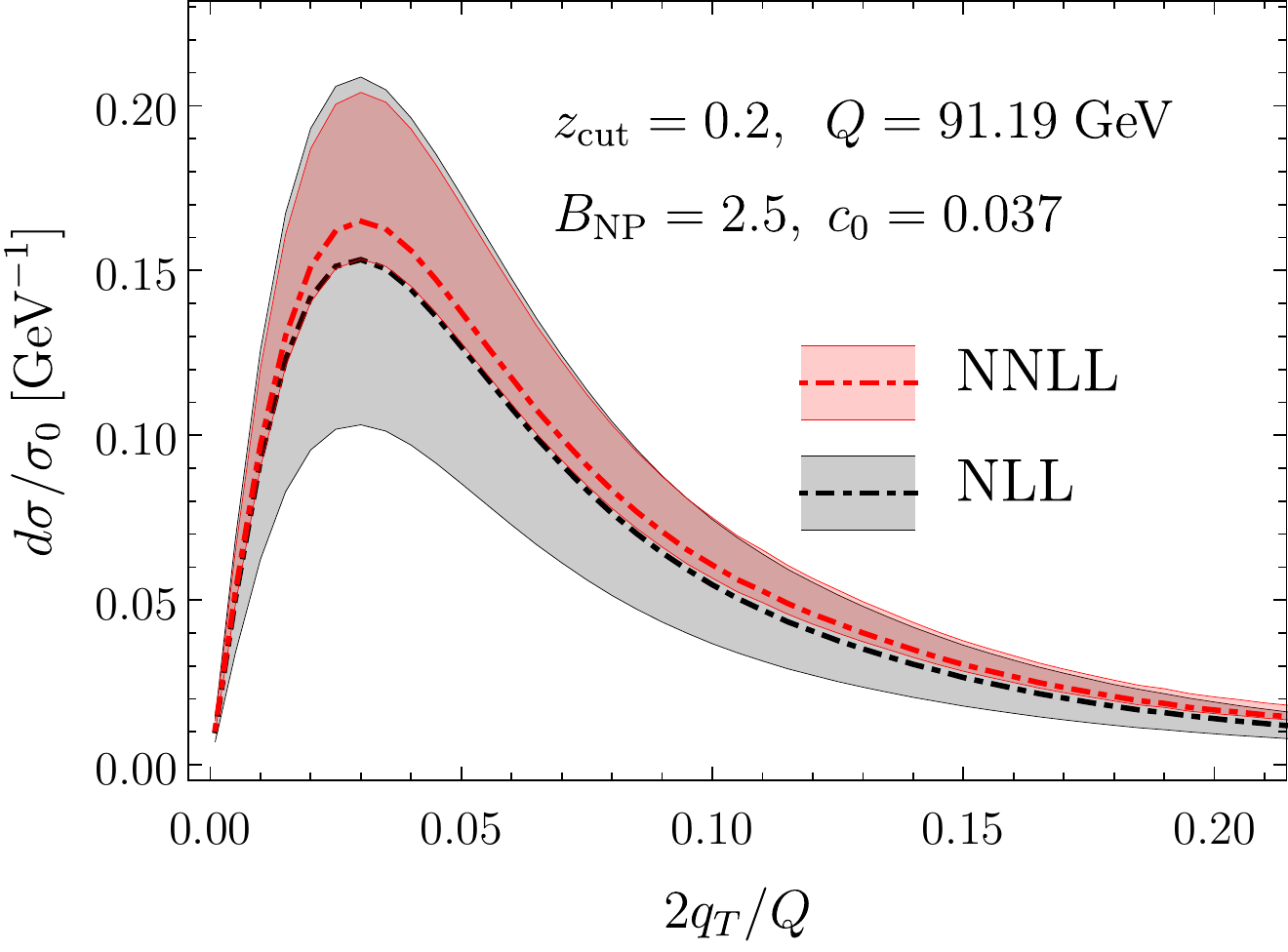}}
  \caption{Transverse momentum de-correlation for $e^+e^-\rightarrow$ dijets with center of mass energy at the Z mass. }
  \label{fig:e+e-Zmass}
\end{figure}

In fig.~\ref{fig:e+e-Zmass} we give the NNLL results including a theoretical uncertainty band. We compare against the NLL cross section and although the error bands seem to be larger than what is typically expected  we can clearly see that the result convergences and the theory error decreases by approximately factor of two. To estimate the theoretical uncertainty we first vary all the factorization scales of a factor 2 (0.5) around their canonical value, then we separately take the envelope of the variations involved in rapidity evolution, $\mu,\mu_{sc}$, and of the ones involved only in the virtuality evolution of the jet function, $\mu_{cs},\mu_J$. The final error bands we show are the quadrature of the two contributions. The reason for this prescription is that rapidity and virtuality evolutions are in principle uncorrelated. The  uncertainty is somewhat larger than what one might expect for a NNLL calculation, and is practically dominated by the variations in the jet function. This is attributed to the small values of the collinear-soft scale, $\mu_{cs}\sim Q\sqrt{\ecut \zcut}$, which approaches the non-perturbative regime even for values of $Q \sim m_Z$. One might attempt to reduce the uncertainty by increasing either $\ecut$ or $\zcut$, but caution is needed not to invalidate the corresponding hierarchy. We will see later that when only the mass of one jet is measured (e.g., in DIS or hadron-jet decorrelation) then the error band decreases significantly.

\section{Jets in DIS}
\label{DIS-jet}

The advent  of new colliders like EIC and LHeC makes the measurement of jets interesting also in semi-inclusive deep inelastic scattering (SIDIS) experiments.
Actually  we want to explore the possibility of using jets to study the TMDPDF.

For the present case we demand that the hard scattering of the lepton on the proton produces a single jet. 
In the Breit frame we measure the transverse component, $\pmb{q}_{T}$, of  the transferred momentum, $q^{\mu} = {k'}^{\mu}- {k}^{\mu}$ with respect to the single groomed jet.  As before, we impose a jet mass cut-off $\ecut$ and the grooming parameter $\zcut$. In this framework the initial state proton is moving along the $-z$ direction and the final state jet is moving in the opposite $+z$-direction, so that we can assign the directions $n$ and $\bar{n}$ to the beam and jet definition. The contribution to this transverse momentum measurement  comes from the initial state radiation which forms part of the TMDPDF and the radiation that fails soft-drop in the final state jet. We demand that there is a single energetic jet with $E_J \sim Q/2 =\sqrt{-q^2} /2$ with accompanying soft radiation.

It is instructive to setup some of the notation that we are using for describing the kinematics in the Breit frame. The virtual photon is assumed to be completely space-like and it has only the $z$ component of the momentum. Defining our light-cone directions $n^{\mu} =(1, 0, 0, +1)$ and $\bar{n}^{\mu} = (1,0,0,-1)$, the resulting photon momentum $q^{\mu}$ is
\begin{equation}
  q^{\mu} = \frac{Q}{2} (n^{\mu} - \bar{n}^{\mu})\ ,
\end{equation}
where $Q^2 = - q^2$ is a positive quantity.  We assume that at the partonic level, a single quark carrying $x$ fraction of the proton longitudinal momentum  undergoes a hard interaction with the virtual photon. 
In this frame, the proton is moving along the $-z$ direction and its momentum can be written as:
\begin{equation}
  P^{\mu} = \frac{Q}{2 x} \bar{n}^{\mu}\ .
\end{equation}
At tree level and by momentum conservation the final state parton will carry  momentum 
\begin{equation}
  xP^\mu + q^\mu = \frac{Q}{2}(n^{\mu} - \bar{n}^{\mu})+ \frac{Q}{2}\bar{n}^{\mu}= \frac{Q}{2} n^{\mu}\;,
\end{equation}
which is exactly opposite in direction to the incoming beam. Of course this will be modified beyond tree-level when initial and final state radiation is included.

\subsection{Schematics for factorization}

Since we are  working with two back-to-back directions,  our usual definition of the soft function holds: in other words the change  from future pointing to past pointing Wilson lines  does not affect its value~\cite{Collins:2011zzd,GarciaEchevarria:2011rb,Echevarria:2014rua,Gaunt:2014ska,Vladimirov:2017ksc}.

Since we still impose the same jet mass measurement on the final state jet,   we have all the modes that we had  in the $e^+e^-$ case.  The main  difference is that  now the initial hadronic state is a TMDPDF.
The form of the factorized cross section follows again  the hierarchy   $Q \gg Q \zcut \gg q_{T}, \; R\sim 1$ and 
\begin{equation}
  \label{eq:fact_ep}
  \dsdqdx = \mathcal{N}(x,Q) H_2(Q,\mu) \times S(\pmb{q}_T) \otimes B_{i\leftarrow h}(x,Q,\pmb{q}_T) \otimes \mathcal{J}_{j}^\perp(\ecut, Q,\zcut, \pmb{q}_T)\ ,
\end{equation}
where $ x = - q^2/(2P \cdot k)$, $k$ is the momentum of the incoming electron, and $\mathcal{N}(x, Q)$ is the over-all normalization which we give later in this section. The un-subtracted TMDPDF is  $B_{i\leftarrow h}$. 
In our rapidity regularization scheme the (subtracted) TMDPDF is defined as 
\begin{align}
  F_{i\leftarrow h}(x,\boldsymbol b; \mu, \zeta)=\sqrt{S(\pmb{b})}B_{i\leftarrow h}(x,Q,\pmb{b}) .
\end{align}
At perturbative values of $q_{T}$, the  $F_{i\leftarrow h}$ can be matched onto the collinear PDF. The matching coefficients at NNLO are evaluated in~\cite{Gehrmann:2014yya,Echevarria:2016scs}
and in the appendix we review  some one-loop results.
Once the subtracted quantities are included  we can write
\begin{equation}
  \label{eq:fact_ep_II}
  \dsdqdx = \mathcal{N}(x,Q) H_2(Q,\mu) \int \frac{d \pmb{b}}{4\pi^2} e^{i \pmb{b} \cdot \pmb{q}_{T} }  F_{i\leftarrow h}(x,Q,\pmb{b},\mu,\zeta_A) \mathcal{J}_{j}^\perp(\ecut, Q,\zcut, \pmb{b};\mu,\zeta_B)\ .
\end{equation}
The evolution under renormalization group equations for the TMDPDF  is widely known (see e.g.~\cite{Scimemi:2018xaf, DAlesio:2014mrz,Echevarria:2012pw,Chiu:2011qc}) and we recall a few characteristics here.  One has
\begin{align}
  &\mu\frac{d}{d\mu} F_{f\leftarrow f'}(x,\boldsymbol b, \mu, \zeta)=\gamma_F^f(\mu,\zeta) F_{f\leftarrow f'}(x,\boldsymbol b, \mu, \zeta),\nn\\
  &\zeta \frac{d}{d\zeta} F_{f\leftarrow f'}(x,\boldsymbol b, \mu, \zeta)=-\mathcal{D}^f(\mu,\boldsymbol b) F_{f\leftarrow f'}(x,\boldsymbol b, \mu, \zeta),
\end{align}
where $\mathcal{D}_f$ and $\gamma_F^f$ are the rapidity and UV anomalous dimensions, respectively. 
The integrability  requirement of this couple of equation results in 
\begin{align}
  \mu \frac{d}{d\mu} \left(-\mathcal{D}^f(\mu,\boldsymbol b)\right)=\zeta \frac{d}{d\zeta} \gamma^f_F(\mu,\zeta)=-\Gamma^{\rm cusp}_f
\end{align}
where $\Gamma^{\rm cusp}_f$ is the cusp anomalous dimension. The UV anomalous dimension is written in these terms as
\begin{align}
  \gamma_F^f=\Gamma^{\rm cusp}_f\textbf{l}_\zeta-\gamma_V^f,
\end{align}
$\gamma_V^f$ being the non-cusp part of the anomalous dimension and $\textbf{l}_\zeta=\ln \left(\mu^2/\zeta\right)$. The $\gamma_V$ and $\mathcal{D}$ anomalous dimensions are known up to $\mathcal{O}(a_s^3)$ \cite{Moch:2004pa,Moch:2005tm,Baikov:2009bg,Vladimirov:2016dll,Li:2016ctv}. A numerical calculation for the four-loop cusp anomalous dimension was recently given in \cite{Vogt:2018miu}. All the evolution equations are the same for the  case of TMD fragmentation functions, and we do not discuss them any more here.



\subsection{Derivation of the factorized cross section using jets}
In this section we provide some details for the factorization of the SIDIS cross section in eq.~(\ref{eq:fact_ep}, \ref{eq:fact_ep_II}).
The scattering amplitude for the process $e p \to e f$ where $f$ is the final state is given by:
\begin{equation}
  i M(ep \to ef ) = (-ie^2)\bar{u}(k') \gamma_\mu u(k) \frac{1}{q^2} \Langle f \Bvert J^{\mu}(0) \Bvert p(P) \Rangle\;,
\end{equation}
and thus the corresponding cross section is given by
\begin{multline}
  d\sigma (ep \to ef) = \frac{e^4}{4(s-m^2)} \int \frac{d^3 k'}{2 (2\pi)^3 E_{k'}} \; \text{tr}\lb \slashed{k}\gamma_{\mu} \slashed{k'} \gamma_{\nu}\rb \\
  \sum_f \int d \Pi_f \;\Langle p(P) \Bvert J^{\dag \mu}(0) \Bvert f \Rangle \Langle f \Bvert J^{\nu}(0) \Bvert p(P) \Rangle (2\pi)^4 \delta^{(4)} (q + P -p_f)\;,
\end{multline}
where $q= k'-k$. We can use the standard parametrization of the final electron phase-space to write:
\begin{equation}
  \int \frac{d^3 k'}{2 (2\pi)^3 E_{k'}}  = dxdy \frac{y s}{(4\pi)^2}\;,
\end{equation}
where $y = (2P \cdot q)/(2P \cdot k)$ and $s$ is the hadronic Mandelstam variable. We then get,
\begin{equation}
  \frac{d\sigma}{dxdy} (ep \to ef) = L_{\mu\nu}(k,k') \sum_f \int d^4 r e^{i q \cdot r}\int d \Pi_f \;\Langle p(P) \Bvert J^{\dag \mu}(0) \Bvert f \Rangle \Langle f \Bvert J^{\nu}(x) \Bvert p(P) \Rangle\;,
\end{equation}
where   $r^{\mu}$ is Fourier conjugate of the momenta $q^{\mu}$ and  $L^{\mu\nu}$ is the leptonic tensor,
\begin{equation}
  L_{\mu\nu}(k,k') \equiv \frac{\alpha^2 y s}{4(s-m^2)} \text{tr}\lb \slashed{k}\gamma_{\mu} \slashed{k'} \gamma_{\nu}\rb\;.
\end{equation}
The next step  is to project the hadronic final state $\vert f \rangle $ onto the one that corresponds to the measurement that we are proposing, i.e.,
\begin{equation}
  \int d\Pi_f \Bvert f \Rangle \Langle f \Bvert\;\; \to \int d\pmb{q}_{T} z dz \int d\Pi_{f[\text{g-jet}(z \pmb{q}_{T}, z)]} \Bvert f \Rangle \Langle f \Bvert\;.
\end{equation}
We can now match the full theory hadronic current $J_{\mu}(x)$ onto the SCET$_+$~\cite{Bauer:2011uc} current working in the Breit frame, 
\begin{equation}
  J^{\mu}(x) = C^{\mu \nu}(Q) \lb \bar{\chi}_{n,Q}  S_{n}^{\dag} W_t^{\dag} U_n\gamma_{\nu}  S_{\bar{n}}\chi_{\bar{n},Q}  \rb + \mathcal{O}(\lambda)\;,
\end{equation}
where $\lambda$ is the power counting parameter of our EFT which will turn out to be $q_{\perp}/Q \sim \ecut/Q$. Note that in the same step, through BPS field redefinition, we decoupled the  collinear soft modes from the collinear modes and hence the presence of the $U_n$ Wilson lines. In the matching we also have the soft Wilson lines $S_n$. From the kinematic constraints of the measurement and since all the modes that are present in the projected final state are decoupled from each other at the level of the Lagrangian, (we assume that contributions from Glauber gluon exchanges cancel) it is possible to factorize the final state as follows,
\begin{equation}
  \Bvert f \Rangle  \;\;\to\;\;  \Bvert X_{\bar{n}} \Rangle  \Bvert X_{n} \Rangle   \Bvert X_{s} \Rangle  \Bvert X_{sc} \Rangle \;,
\end{equation}
where we have included in $X_n$ all possible modes that contribute to the invariant mass measurement. Refactorization of the $n-$collinear sector follows from the same steps as in the case of electron-positron annihilation presented in ref.~\cite{Frye:2016aiz}. We are now ready to factorize the cross section into individual SCET matrix elements. In the final result one needs to be careful regarding all the index contractions and the tensor structures. This was carefully considered in ref.~\cite{Kang:2013nha}. In addition we are considering the case where the frame we are working is rotated such that the transverse momentum of the groomed jet is zero. After all rearrangements we get,
\begin{multline}
  \frac{d\sigma}{dxdydzd\pmb{q}_{T}} (ep \to ef) = \sigma_0 (x,Q) \times H_2 (Q)  \int d^4r e^{i q \cdot r} \\
  \frac{1}{N_c}\sum_{X_s} \Langle 0 \Bvert S_{n} S_{\bar{n}}^{\dag}(r_{\perp}) \Bvert X_s \Rangle \Langle  X_s\Bvert  S_{\bar{n}} S_{n}^{\dag}(0) \Bvert 0 \Rangle \\
  \times  \sum_{X_{\bar{n}}} \Langle p(P) \Bvert \bar{\chi}_{\bar{n}}(r^{+},r_{\perp}) \frac{\gamma^{+}}{2} \Bvert X_{\bar{n}} \Rangle \Langle X_{\bar{n}} \Bvert \chi_{\bar{n}}(0) \Bvert p(P)  \Rangle \\
  \times  \frac{1}{N_c} \sum_{X_{sc}} \Langle 0 \Bvert U_n^{\dag} W_t (r_{\perp}) \Bvert X_{sc} \Rangle \Langle  X_{sc}\Bvert W_t^{\dag} U_n(0) \Bvert 0 \Rangle \\
  \times \frac{z}{2N_c}  \text{tr}\sum_{X_{n}} \Langle 0 \Bvert   \frac{\gamma^{-}}{2} \chi_{n}(r^{-},r_{\perp}) \Bvert X_{n} \Rangle \Langle X_{n} \Bvert \bar{\chi}_{n}(0) \Bvert 0  \Rangle \Bvert_{p_{\perp}^{X_n} = 0}\;.
\end{multline}
The hard matching coefficient in general has two Lorentz structures, given the two types of currents, vector and axial. For the case of photon with vector current, we simply have $H^{\mu\nu} \sim g^{\mu\nu}_{\perp}$. We have also multipole-expanded the final result. To proceed with the factorization theorem in momentum space, we remove $r_\perp$ dependence  from the various EFT matrix elements  by acting the corresponding fields on the final states. This gives us
\begin{multline}
  \frac{d\sigma}{dxdydzd\pmb{q}_{T}} (ep \to ef) = \sigma_0 (x,Q) \times H_2 (Q) \int d^4r e^{i q \cdot r+i(\pmb{p}^{X^{R}_{\bar{n}}}_{\perp} + \pmb{p}_{\perp}^{S}) \cdot \pmb{r}_{\perp}} \\
  \frac{1}{N_c}  \sum_{X_s} \Langle 0 \Bvert S_{n} S_{\bar{n}}^{\dag}(0) \Bvert X_s \Rangle \Langle  X_s\Bvert  S_{\bar{n}} S_{n}^{\dag}(0) \Bvert 0 \Rangle \\
  \times  \sum_{X_{\bar{n}}} \Langle p(P) \Bvert \bar{\chi}_{\bar{n}}(r^{+}, 0_{\perp}) \frac{\gamma^{+}}{2} \Bvert X_{\bar{n}} \Rangle \Langle X_{\bar{n}} \Bvert \chi_{\bar{n}}(0) \Bvert p(P)  \Rangle \\
  \times\frac{1}{N_c}  \sum_{X_{sc}} \Langle 0 \Bvert U_n^{\dag} W_t (0) \Bvert X_{sc} \Rangle \Langle  X_{sc}\Bvert W_t^{\dag} U_n(0) \Bvert 0 \Rangle \\
  \times \frac{z}{2N_{c}} \; \text{tr}\sum_{X_{n}} \Langle 0 \Bvert   \frac{\gamma^{-}}{2} \chi_{n}(r^{-},0_{\perp}) \Bvert X_{n} \Rangle \Langle X_{n} \Bvert \bar{\chi}_{n}(0) \Bvert 0 \Rangle \Bvert_{p_{\perp}^{X_n} = 0}\;,
\end{multline}
where
\begin{equation}
  p_{\perp}^{X_{\bar{n}}^{R}} \Bvert_{p_{\perp}^{\text{g-jet}} = 0}  =  p_{\perp}^{X_{\bar{n}}} - P_{\perp} \Bvert_{p_{\perp}^{\text{g-jet}} = 0} = p_{\perp}^{X_{\bar{n}}} \Bvert_{P_{\perp} = 0} \lp 1 + \mathcal{O}(\lambda)\rp\;,
\end{equation}
is the difference in the transverse momentum of the recoiling initial state collinear radiation and the proton with respect to the hadrons direction, which up to power-corrections of order $\mathcal{O}(\lambda)$ is simply the transverse momentum of the recoiling radiation with respect to the proton. Performing the integral over $d^4r$ we get:
\begin{multline}
  \frac{d\sigma}{dxdydzd\pmb{q}_{T}} (ep \to ef) = \sigma_0 (x,Q) \times H_2 (Q) \delta^{(2)} (\pmb{q}_{T}+ \pmb{p}^{X^{R}_{\bar{n}}}_{\perp} + \pmb{p}_{\perp}^{X_s} + \pmb{p}_{\perp}^{X_{sc}} ) \\
  \frac{1}{N_c}  \sum_{X_s} \Langle 0 \Bvert S_{n} S_{\bar{n}}^{\dag}(0) \Bvert X_s \Rangle \Langle  X_s\Bvert  S_{\bar{n}} S_{n}^{\dag}(0) \Bvert 0 \Rangle \\
  \times  \sum_{X_{\bar{n}}} \Langle p(P) \Bvert \bar{\chi}_{\bar{n}}(0) \frac{\gamma^{+}}{2} \delta(q^{-} - p_{X_{\bar{n}}}^{-})\Bvert X_{\bar{n}} \Rangle \Langle X_{\bar{n}} \Bvert \chi_{\bar{n}}(0) \Bvert p(P)  \Rangle \\
  \times \frac{1}{N_c}  \sum_{X_{sc}} \Langle 0 \Bvert U_n^{\dag} W_t (0) \Bvert X_{sc} \Rangle \Langle  X_{sc}\Bvert W_t^{\dag} U_n(0) \Bvert 0 \Rangle \\
  \times \frac{z}{2N_{c}} \; \text{tr}\sum_{X_{n}} \Langle 0 \Bvert   \frac{\gamma^{-}}{2} \chi_{n}(0)  \delta(q^{+} - p_{ X_{n}}^{+}) \Bvert X_{n} \Rangle \Langle X_{n} \Bvert \bar{\chi}_{n}(0) \Bvert 0  \Rangle \Bvert_{p_{\perp}^{X_n} = 0}\;.
\end{multline}
In order to simplify our result further we introduce ``measurement'' delta functions for the soft and initial state matrix elements. This will allow us to absorb the $p_{\perp}^{X_i}$ into the corresponding matrix elements and use 

\begin{equation}
  \bmat{1}_i = \sum_{X_i} \Bvert X_i \Rangle \Langle X_i \Bvert\;,
\end{equation}
to further simplify the form of EFT matrix elements. We also perform a type-I RPI transformation in order to rewrite the proton matrix elements as function of fields with respect to the initial state proton axis. We thus get
\begin{multline}
  \frac{d\sigma}{dxdydzd\pmb{q}_{T}} (ep \to ef) = \sigma_0 (x,Q) \times H_2 (Q) \int d\pmb{p}_{\perp}^{\;s} d\pmb{p}_{\perp}^{\;sc} d\pmb{p}_{\perp}^{\;c} \;\;\delta^{(2)} ( \pmb{q}_{T}+\pmb{p}^{c}_{\perp} + \pmb{p}_{\perp}^{s} +\pmb{p}_{\perp}^{sc} ) \\
  \frac{1}{N_c}  \Langle 0 \Bvert T\lp S_{n} S_{\bar{n}}^{\dag}(0)\rp \delta^{(2)}(\pmb{p}_{\perp}^{s} - \pmb{\mathcal{P}}_{\perp}) \bar{T} \lp S_{\bar{n}} S_{n}^{\dag}(0) \rp \Bvert 0 \Rangle \\
  \times  \Langle p(P) \Bvert \bar{\chi}_{\bar{n}}(0) \frac{\gamma^{+}}{2} \delta(q^{-} - \mathcal{P}^{-})  \delta^{(2)}(\pmb{p}_{\perp}^{c} - \pmb{\mathcal{P}}_{\perp})  \chi_{\bar{n}}(0) \Bvert p(P)  \Rangle \Bvert_{P_{\perp} = 0} \\
  \times \frac{1}{N_c}   \Langle 0 \Bvert  T\lp U_n^{\dag} W_t (0) \rp \mathcal{M}_\perp^{SD}  \bar{T} \lp W_t^{\dag} U_n(0) \rp \Bvert 0 \Rangle \\
  \times \frac{z}{2N_{c}} \; \text{tr}\sum_{X_{n}} \Langle 0 \Bvert   \frac{\gamma^{-}}{2} \chi_{n}(0)  \delta(q^{+} - \mathcal{P}^{+}) \Bvert X_{n} \Rangle \Langle X_{n} \Bvert \bar{\chi}_{n}(0) \Bvert 0  \Rangle \Bvert_{p_{\perp}^{X_{n}} = 0}\;,
\end{multline}
where $\mathcal{M}_\perp^{SD}$ is the measurement function given in eq.~(\ref{eq:meas_sc}). Since we are considering only large radius jets with $R\gtrsim 1$ we may trivially perform the integration of the energy fraction $z$ using $Q\simeq p^{X_n}_+$ up to power corrections. Also performing change of integration variables,
\begin{equation}
  dx dy = \frac{dx dQ^2}{xs} \;, 
\end{equation}
we get eq.~(\ref{eq:fact_ep}) with
\begin{equation}
  \mathcal{N}(x,Q) = \frac{\sigma_{0}(x,Q)}{xs}\;,
\end{equation}
and the matrix elements involved in the functions $S$, $B$, and $\mathcal{J}$ are given in the appendix. For the case of groomed jets with invariant mass cutoff it is possible to refactorize the jet function.  This is done in ref.~\cite{Frye:2016aiz} and thus we do not demonstrate it here. Then integrating over $e \in (0,\ecut)$  gives the dependence of the jet function in the parameter $\ecut$. This is identical to the analysis in the previous section on $e^+e^-$. This is our final result for the factorization theorem in DIS.



\subsection{Numerical results for DIS}
\label{sec:numerics-DIS}

\begin{figure}[h!]
  \centerline{\includegraphics[width =  \textwidth]{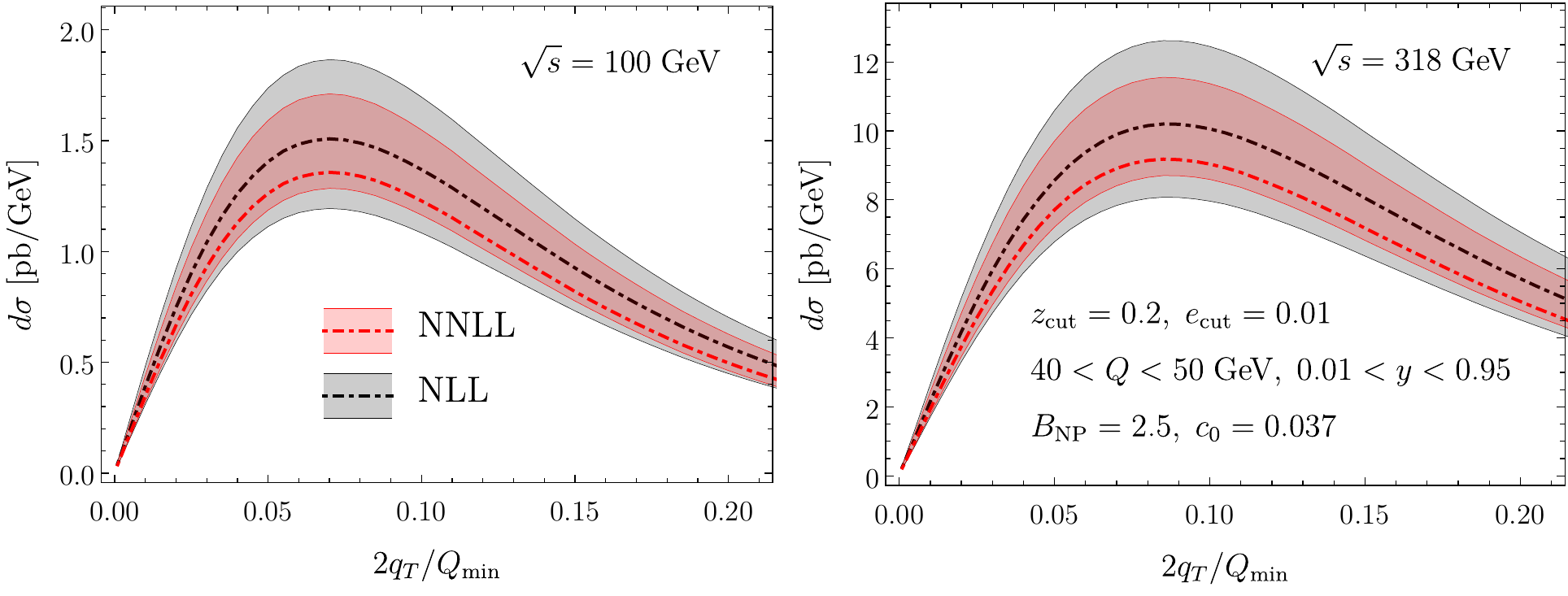}}
  \caption{The NLL and NNLL  TMD spectra for groomed jets in DIS for EIC (left: $\sqrt{100}$ GeV) and HERA (right: $\sqrt{s} = 318$ GeV) kinematics.  The cross section are integrated in $y = Q^2/(xs)$ and $Q = \sqrt{-q^2}$ (see details in the main text). }
  \label{fig:scales}
\end{figure}

In this section we use the factorization theorem in eq.~(\ref{eq:fact_ep}) to obtain numerical results for the TMD spectrum of groomed jets in DIS process. Our analysis is done for two center-of-mass energies, EIC: $\sqrt{s} = 100$ GeV and HERA: $318$ GeV.  For both energies we integrate over $y = Q^2/(xs) $ and $Q = \sqrt{-q^2}$ in the regions $0.01< y<0.95$ and $40< Q<50$ GeV. For the TMDPDFs we use the fits obtained from Drell-Yan data~\cite{Bertone:2019nxa} with the use of $\zeta$-prescription.  
In fig.~\ref{fig:scales} we show our results for NLL and NNLL accuracies for the two center of mass choices, including theoretical uncertainties. We estimate the theoretical  scale variations  as described in  sec.~\ref{sec:numerics-ee}. The groomed jet parameters that we choose are the same as in the di-lepton case: $\beta =0$, $\zcut =0.2$, and $\ecut = 0.01$. As before we find good convergence between the  NLL and NNLL result.  The absolute value of  theoretical scale variation is improvable  with higher logarithmic accuracy (NNLL-prime or perhaps N$^3$LL), which needs the explicit calculation of several  jet hadronic matrix elements at two loops. 

\begin{figure}[t!]
  \centerline{\includegraphics[width =  \textwidth]{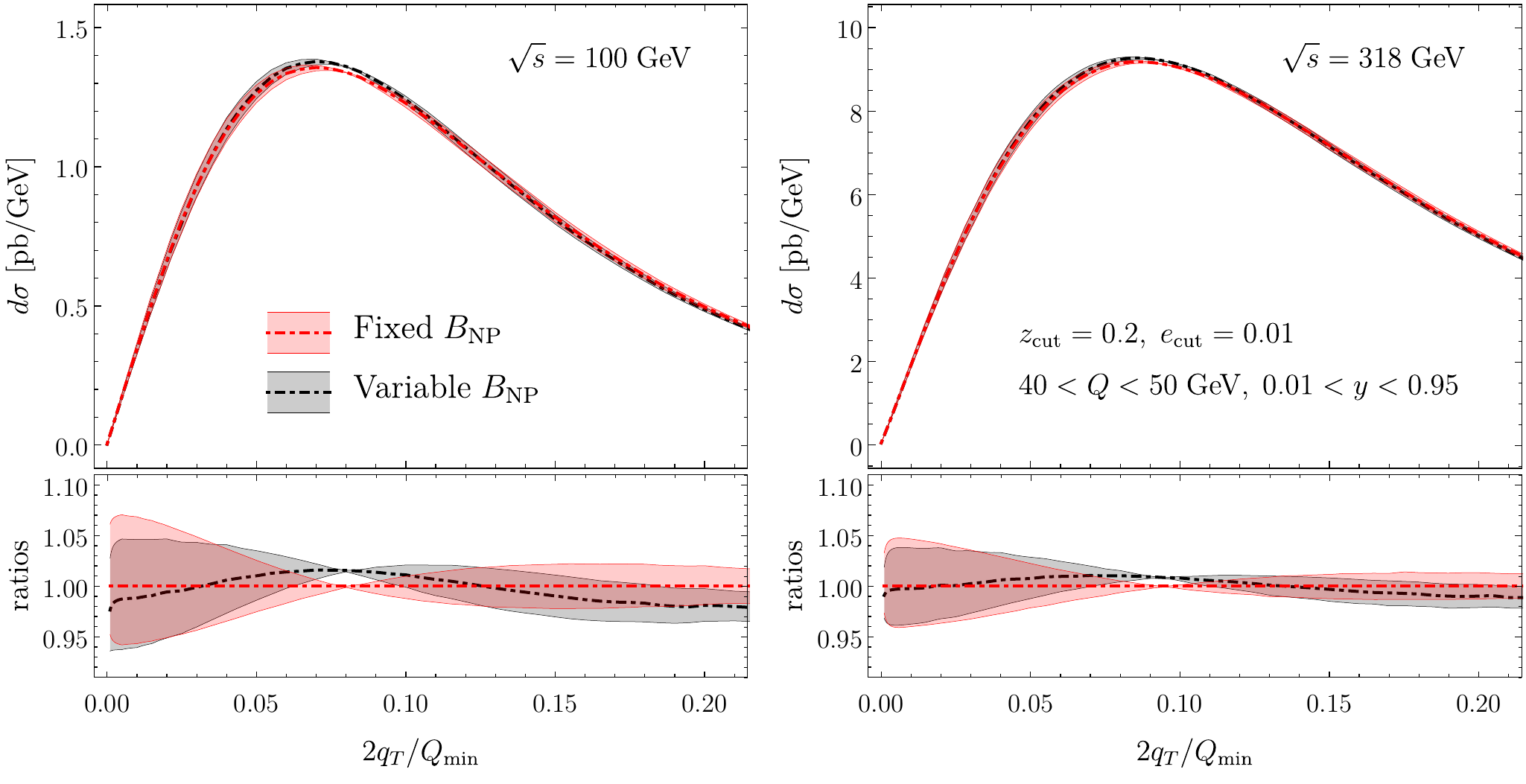}}
  \caption{The NNLL cross-section including modeling of the initial hadronic state effects fitted from Derll-Yan processes using two different scenes: fixed and variable $B_{\text{NP}}$.}
  \label{fig:NP}
\end{figure}

We further investigate the size of the uncertainty due to the hadronic initial state and the non-perturbative effects induced by TMD evolution. We do that by varying the model parameters as constrained by the phenomenological analysis in ref. ~\cite{Bertone:2019nxa}  for our NNLL result. The results are shown in figure~\ref{fig:NP}.  We consider both variable and fixed $B_{\text{NP}}=2.5$ GeV$^{-1}$  (for details on the difference  of the two schemes see~\cite{Bertone:2019nxa}).  We find that the effects (for our kinematics) are particularly small, of the order of $\sim 5\%$, which is much smaller than the theoretical uncertainties. This suggests that we need a better control over the theoretical uncertainties in order to further constrain TMD distributions from groomed jets in DIS.  As mentioned earlier the  uncertainty can be mitigated with higher logarithmic accuracy or by choosing larger values of $\ecut$, still compatible with factorization. This, will require to treat the region III shown in fig.~\ref{fig:regions}. For this reason it  is interesting  to investigate the range of values of $\ecut$ for which the  energetic wide angle radiation is avoided.
\section{Hadronization effects }
\label{hadronization}
One of the  goals of the paper is to study the non-perturbative effects associated with TMD distributions, in this case the TMDPDF. Usually in any experiment, there are multiple sources of non-perturbative corrections associated with both the initial and final states. To have access to a specific source of corrections, its therefore necessary to separate out the pieces of interest from the uninteresting ones, which in this case constitute the final state hadronization corrections. To access the TMD then, we must already have a good extraction of the rest of the non-perturbative effects. This is the reason  why we consider distinct experiments in this paper. The idea, as we shall demonstrate, is that the final state hadronization corrections are exactly the same in the two experiments. The $e^+e^- \to \dijets$  case can be used to extract out all the final state hadronization corrections, which can then be used for DIS.

For the $e^+e^-$ observable, the factorization takes the form in eq.~(\ref{eq:fact-IIa}), we can then study the non-perturbative corrections for each collinear object $\mathcal{J}^{\perp}_{i}$, which by symmetry, are the same for the two objects.  If we now look at the factorization for DIS, eq.~(\ref{eq:fact_ep_II}), the key point to note is that $\mathcal{J}_{j}^\perp(\ecut, Q,\zcut, \pmb{b};\mu,\zeta_B)$ is the same object that appears in the the case of $e^+e^-$, while $ F_{i\leftarrow h} $ is just the TMDPDF. Thus it now becomes possible to exclusively access the complete TMDPDF. We now wish to systematically list the sources of the  non-perturbative corrections associated with each factorized function that appear in our cross section.

In order to use  jets it is important to consider all the non-perturbative effects for the case of our observables and in particular the ones coming from the implementation of (groomed) jets.  In fig.~\ref{fig:hadronization} we have shown that  such corrections are expected to be particularly small and we provide here a discussion about their origin from a theory perspective. We have two measurements on the jet:  the  jet mass, which is ultimately integrated over  some interval  and acts as a normalization, and   the transverse momentum ($p_{\perp}$) of the radiation that is groomed away. Since we are interested in the shape of the $q_T$ spectrum, we will only consider the non-perturbative effects in cross sections sensitive to it. As was explained in sec.~\ref{sec:main-1}, we are working in the region II of EFT and we are going to  discuss how non-perturbative effect arise when we increase the value of $q_T$ (that is, we discuss here the non-perturbative corrections in the small-$b$ limit, where $b \equiv \vert \pmb{b} \vert$).   Our factorization theorem  has four functions in the IR, the collinear, the global soft, the collinear-soft, the soft-collinear functions, see eq.~(\ref{eq:jet-fact-II}-\ref{eq:jj}),  and all of them can potentially contribute to non-perturbative power corrections. Even though the collinear and collinear-soft functions do not contribute to $\pmb{q}_T$ perturbatively, they can still give a non-perturbative power correction to the $\bmat{q}_T$ spectrum\footnote{There are also power corrections of similar magnitude in this region due to the factorization of the $sc$ function from the $cs$, but they are perturbative in nature and can be handled by making a smooth transition to region III.}.

There are two types of non-perturbative corrections that we will consider here. We call \emph{shift}  non-perturbative effects the ones which are not altered by the pass and fail procedure of the grooming conditions. An example is  the global soft function that is independent of the grooming procedure and  it is common to other TMD analysis. We refer to this kind of correction as \emph{shift}  non-perturbative effects since, as we will see later, in the simplest case it generates a shift in the TMD spectrum. The second correction instead is related to the grooming procedure  with $cs$ and $sc$  soft functions and  the jet shape function. In this case non-perturbative effects are driven by the so called ``non-perturbative particles'' and it is obviously only possible when perturbative modes are on the boundary of passing and failing soft-drop. We refer to these contributions as \emph{boundary} non-perturbative effects. 


\subsection{\emph{Shift} non-perturbative correction}
For the case of shift correction, we assume that the soft-drop condition remains unaltered by any non-perturbative emissions. Now consider the contribution to the shift correction by each function in turn. 

The non-perturbative part of the global soft function defined in eq.~(\ref{eq:soft}) has been studied in the literature in several frameworks \cite{Beneke:1995pq,Korchemsky:1994is,Korchemsky:1997sy,Beneke:1997sr,Becher:2013iya,Scimemi:2016ffw}. Up to ${\cal O}(b^4)$ terms it can be written as 
\bea
\langle 0| T[S_nS^{\dagger}_{\bar n} (\pmb{b})] \bar{T}[ S_{\bar n}S_{n}^{\dagger}(0)] \vert 0 \rangle 
= \tilde S( b) +b^2  \; \bar{C}^{(s)}_i( b)  \langle 0| O^{i}|0 \rangle \;,
\eea
where $O^i$ is the complete set of local operators that have the same quantum numbers as the soft function. Summation over $i$ is implied. Here $\tilde{S}$  is the perturbative calculable part of the soft function and it contains rapidity and UV divergences as well as the rest of other terms in the equation. We can pull this out as a common factor to write
\bea
&\langle 0| T[S_nS^{\dagger}_{\bar n} (\pmb{b})] \bar{T}[ S_{\bar n}S_{n}^{\dagger}(0)] |0\rangle
&= \tilde S( b) \lp 1+  b^2\; C^{(s)}_i( b)  \langle0| O^{i}|0\rangle \rp \;.
\eea
To maintain the UV scale invariance  of the cross section, we need that the second term in the brackets  be independent of UV divergences. However  additional rapidity divergences may be present in the non-perturbative matrix element on the r.h.s. that cancel with the  corresponding rapidity divergence arising in the non-perturbative power corrections to the collinear  or soft-collinear functions.  This is related to the origin of the non-perturbative correction to the rapidity anomalous dimension and it  is usually included also in TMD analysis.

We can perform a similar analysis for the soft-collinear ($sc$) function. When an $sc$ (perturbative) mode passes soft-drop, then it does not contribute to $q_T$ since it becomes part of the groomed jet. But since it has a large + component, it drives the  groomed jet mass outside the region of measurement and hence such events are dropped. Therefore, we only need to consider the case when the $sc$ mode fails soft drop. In this case the non-perturbative emission contributes to the $q_T$ measurement if it lies outside the groomed jet. Given the angular scaling of this mode, which is much larger than the collinear-soft ($cs$) and collinear modes that form the groomed jet, the phase space region available is effectively unconstrained (this is also the reason why we ignore any phase space constraints on the soft non-perturbative emissions). Hence the correction in this case will also be a simple shift type and is implemented in the same manner as in the case of the global soft function. As before, we can pull out a common perturbative factor (that includes the perturbative soft drop condition), and write 
\bea
\tilde{S}_{sc}^{\perp}(b,\zcut) \Bvert_{\text{hadr.}}  = \tilde{S}_{sc}^{\perp}( b, Q\zcut ) \lp 1 + b^2 C^{(sc)}_i( b,\zcut) \; \langle 0| O^i |0\rangle \rp .
\eea
Notice that  now all the $z_\cut$ dependence of the power correction is included in the perturbative calculable coefficient $C^{(sc)}( b,\zcut)$, which multiplies the same non-perturbative power correction present also in the global soft function case. The calculation of  $C^{(s)},\; C^{(sc)}$ is doable perturbatively, although this consideration goes beyond the present work.

We can then combine all shift corrections that have an unconstrained phase space for non-perturbative emissions together so that in $b$ space we have a multiplicative correction to the perturbative cross section of the form
\bea 
S  S_{sc}^{\perp} \Bvert_{\text{hadr.}} = (1+b^2 (\Omega_s+\Omega_{sc}) )S S_{sc}^{\perp} \Bvert_{\text{pert.}} \;,
\eea
where $\Omega_s$ is the same as the TMD case and  $\Omega_s$ is a single parameter to be fitted from $e^+e^-$ experiments. It is clear that, in the event of  non-trivial $C^{\{(s),\, (sc)\}}$, $\Omega_{s,\, sc}$ can have a mild (logarithmic) dependence on $q_T$ so that this model will work well over a limited range of $q_T$ which may be sufficient for most cases.

We now consider the shift corrections  coming from the collinear-soft and the collinear functions.  Since these modes determine the region of the groomed jet, we can consider two possible scenarios which give a non-trivial power correction.

\begin{enumerate}
\item{Collinear-soft ($cs$) particles pass soft-drop:}\\
  If the  $cs$ particles pass the  soft-drop (for phase space see figure~\ref{fig:shift}(a)) then any non-perturbative emission scaling as the $cs$ mode can contribute to $q_T$ when it lies outside the groomed jet. In this case, we need to calculate the catchment area of the groomed jet that is determined by the angular distance of the $cs$ subject that passed soft-drop. As was pointed out in \cite{Hoang:2019ceu}, it is possible at NLL, using a coherent branching formalism, to factorize a purely non-perturbative function from all the calculable perturbative effects (including grooming). A detailed analysis of these corrections will be presented in a future work.

  \begin{figure}[h!]
    \centerline{\scalebox{1}{\includegraphics{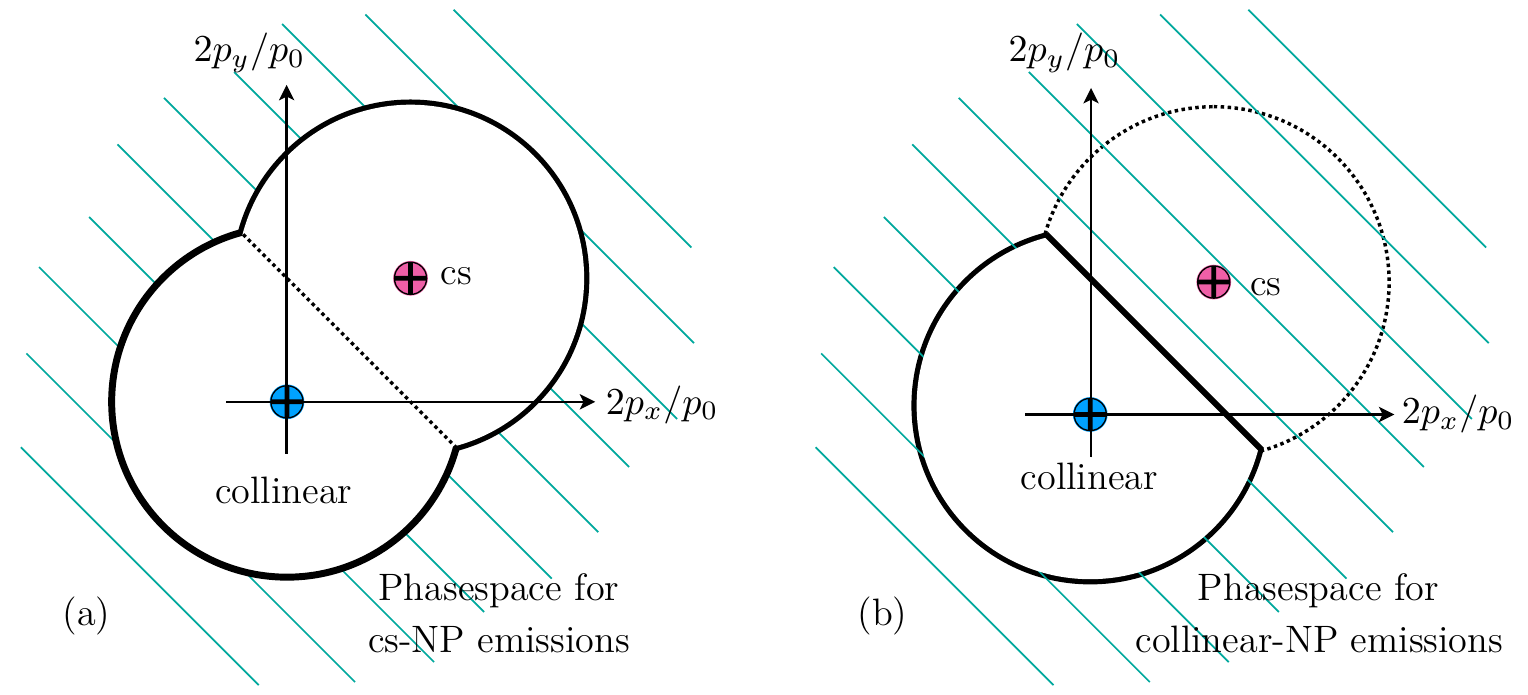}}}
    \caption{(a) When the collinear-soft ($cs$) function passes soft drop, the non-perturbative (NP) emissions, with the angular scaling of the $cs$ mode , with a virtuality $\Lambda_{QCD}$  must fall in the phase space shown by the blue shaded area in order to contribute to $q_T$. (b) When the $cs$ function fails soft drop, the NP emission with the angular scaling of the collinear modes must not be clustered with the collinear sub-jet in order to contribute to $q_T$.}
    \label{fig:shift}
  \end{figure}

\item{Collinear-soft particles fail soft-drop:}\\
  In this case collinear modes are the only ones that pass soft-drop  (for phase space see figure~\ref{fig:shift}(b)), so that any non-perturbative mode scaling as $cs$ has an unconstrained phase space, by the same logic as for the soft and the $sc$ functions, so that we get a simple shift correction of the same form as the soft, $sc$ and TMD collinear functions.\footnote{Technically in this case the perturbative value of $p_{\perp cs}$ would give a larger correction. However, this correction can eventually be handled by transitioning to a new EFT in which the $sc$ and $cs$ functions merge together. For now we will ignore them and only keep track of the other non-perturbative corrections.} There is another possible interesting correction that will come from the collinear NP emission that lies outside the catchment region that is now determined by the collinear modes alone.

  In this case there are two ways of approaching the problem. In one, we consider separating out the non-perturbative corrections before factorizing the $cs$ and collinear modes. The other way is to realize that in the case where $cs$ fails soft-drop, the entire groomed jet mass measurement comes from the jet function alone and using this condition we can define a catchment area for the collinear non-perturbative emissions \textit{without explicitly accessing any information from the $cs$ function}, so that the factorization between the collinear and $cs$ modes is maintained. In this case, we can do a diagrammatic analysis, similar to \cite{Hoang:2019ceu}, for the collinear function, to check if it is possible to factorize the non-perturbative effects from the perturbative. We leave this work for the future.

\end{enumerate}
\subsection{Boundary corrections}
We now consider boundary corrections that leave the $q_T$ measurement function unchanged but only require an expansion of the soft-drop condition in $q^-/Q$.  The functions that do not explicitly have a soft-drop condition can then be ignored, which leaves us with only the $sc$ and $cs$ functions. We can follow the same line of reasoning as in \cite{Hoang:2019ceu}. 

\begin{figure}[h!]
  \centerline{\scalebox{1}{\includegraphics{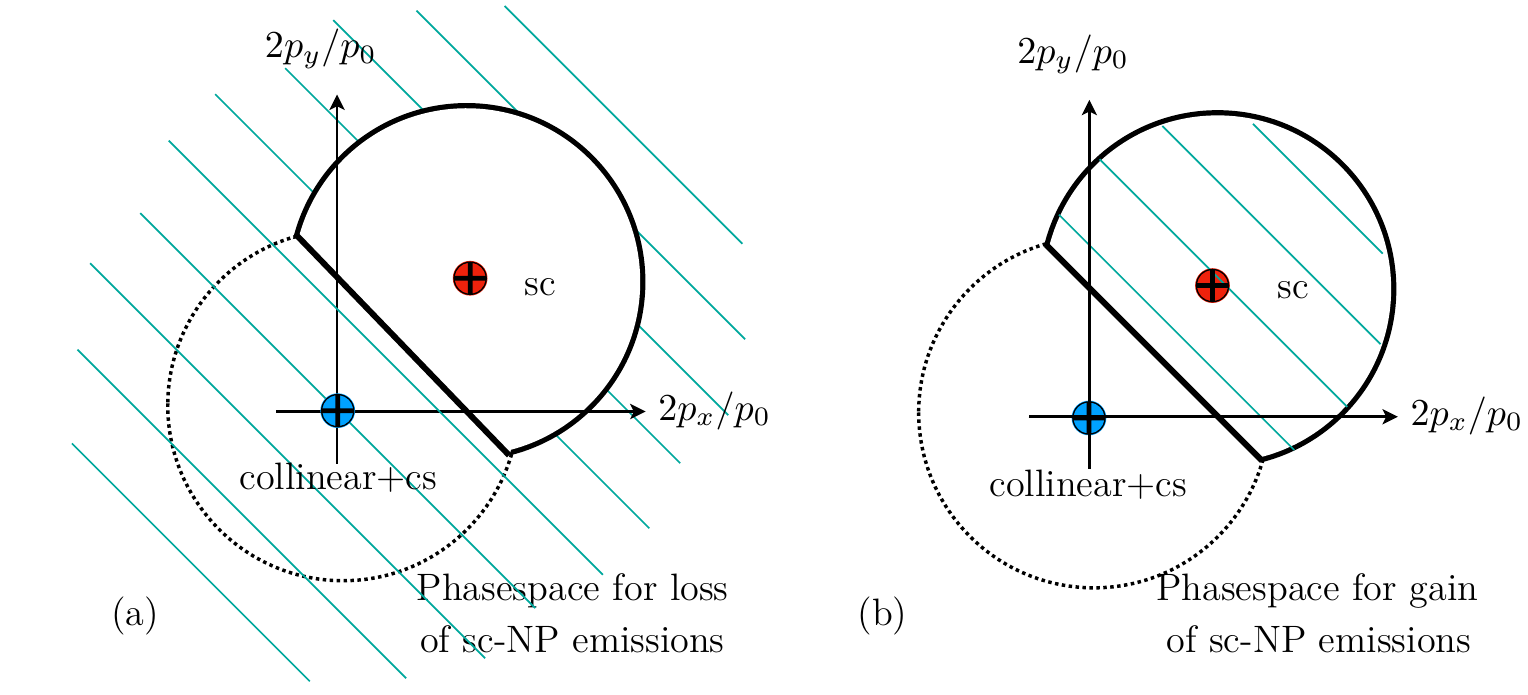}}}
  \caption{(a) The case where the sc subjet loses an NP emission (b) The case when the sc subjet gains an NP emission}
  \label{fig:Boundary}
\end{figure}

\begin{enumerate}
\item{$sc$ emissions}\\
  In this case we demand that either an addition or removal of the non-perturbative emission  cause the soft-collinear function to fail soft-drop. Otherwise it will drive up the jet mass outside the measured range. If we consider a non-perturbative emission $q^{\mu}$ along with a perturbative momentum $p^{\mu}$, then we can expand out the soft-drop condition in the non-perturbative momentum. 
  We can write the complete measurement function as 
  \bea
  \Theta^{p \pm q} =   \Theta\left( \frac{p+q}{E_J}-z_{\text{cut}}\right) \delta^2(\pmb{p}_{\perp sc}- \pmb{p}_{\perp}\mp \pmb{q}_T)\;,
  \eea
  where $p$ is the momentum of the perturbative $sc$ sub-jet while $q^{\mu}$ is the momentum of the non-perturbative emission. The $\pm$ signs indicate whether the perturbative $cs$ subject gains or loses a non-perturbative momentum after hadronization. 
  In the case where the $sc$ sub-jet gains a non-perturbative emission, the measurement expanded to leading order looks like
  \bea
  \Theta^{p +q} \approx  \Theta_{sd}^{p}\delta^2(\pmb{p}_{\perp sc}-\pmb{p}_{\perp})+\frac{q^-}{E_J} \Theta^{\text{b.c.}}( \theta_q,\theta_p, \Delta \phi) \delta^p_{sd}\Big[\delta^2(\pmb{p}_{\perp sc}- \pmb{p}_{\perp})\Big] \;,
  \eea
  with
  \begin{align}
    \Theta^{p}_{sd} &\equiv \Theta\left( \frac{p}{E_J}-z_{\text{cut}}\right)\;, &  \delta^{p}_{sd} &\equiv \delta\left( \frac{p}{E_J}-z_{\text{cut}}\right)\;.
  \end{align}
  In this case, the non-perturbative emission $q^{\mu}$ gets clustered with the $sc$ subject. Note that we have expanded  $q_i$ from the $\pmb{p}_{\perp}$ measurement since we are working at leading order. The phase-space constraint, $\Theta^{\text{b.c.}}$  (see figure~\ref{fig:Boundary}(a)), , gives the condition that ensures $q^{\mu}$ gets clustered with the $sc$ part.

  The second case is when $q^{\mu}$  is emitted off $p^\mu$ but it is not clustered with the $sc$ jet. The short distance condition now acts on $p-q$, which can then be expanded out to give 
  \bea 
  \Theta^{p -q} \approx  \Theta_{sd}^{p}\delta^2(\pmb{p}_{\perp sc}- \pmb{p}_{\perp})-\frac{q^-}{E_J} \bar{\Theta}^{\text{b.c.}}( \theta_q,\theta_p, \Delta \phi) \delta^p_{sd}\Big[\delta^2(\pmb{p}_{\perp sc}- \pmb{p}_{\perp})\Big]\;,
  \eea
  $\bar \Theta^{\text{b.c.}}$  (see figure~\ref{fig:Boundary}(b)), is the phase space region for $q^{\mu}$ so that it falls outside the sc subjet. We can see that the leading power correction scales as $q^-/E_J$, which, given the angular scaling of the $sc$ mode, scales as $q_Tz_{cut}/Q$. Given a typical value of $z_{cut} \sim 0.1$, this factor is then comparable to the $q_T^2/Q^2$ correction that we get from the shift terms.

\item{Soft -Collinear function}\\
  We expect that since perturbatively this function does not contribute to $q_T$, the boundary correction should have no effect on the $q_T$ measurement. 

\end{enumerate}

We now have listed out all the possible NP corrections to the transverse momentum measurement. 


\section{Conclusions}
\label{conclusion}

In this paper, we have presented the computation of the transverse momentum de-correlation observable for fat jets groomed using the Soft-Drop algorithm. We consider two scattering experiments:  $e^+e^- \rightarrow$ di-jets and semi-inclusive DIS. In the former, we measure the transverse momentum imbalance between the two groomed jets. We impose a jet mass constraint on our jets in order to ensure collimated jet configurations. Simulation using PYTHIA show that grooming greatly reduces the impact of underlying events as well as final state hadronization. We show that  the factorization theorem for this observable involves the universal soft function which also appears in the traditional definition of TMDs. We propose that this observable can be used as a probe of the non-perturbative rapidity anomalous dimension, which is a universal parameter for TMD distributions. We prove within our EFT that the cumulant jet mass constraint only adds to the overall normalization of the perturbative cross section and hence does not impact the shape of the transverse momentum distribution although it does contribute to the uncertainty.  We gather or compute all the ingredients necessary to evaluate the cross section to NNLL accuracy and a numerical study for the cases of interest. In the implementation  we have used the \texttt{artemide} code \cite{web,Scimemi:2017etj,Bertone:2019nxa} which contains the most  recent extraction  TMDPDF at higher perturbative orders. As part of the  numerical analysis we have used the $\zeta$-prescription \cite{Scimemi:2018xaf} which allows a  complete disentanglement of non-perturbative effects of rapidity evolution from the rest. An uncertainty analysis gives us an error band of  approximately $\pm$ 10 $\%$. The main ingredient of this error is the perturbative uncertainty which can be systematically improved.  As shown in fig.~\ref{fig:hadronization} the hadronization corrections at low $q_{T}$ are  significantly smaller than the case of a standard jet axis and it is therefore one of the major advantage of using grooming. These effects are expected to be the  same in $e^+e^-$ and SIDIS because of the factorization of the cross section. In the case of $e^+e^-$  these corrections constitute all of non-perturbative effects and they are associated with the final state shower. In order to do a meaningful extraction of non-perturbative parameters in this case, it is therefore necessary to improve the uncertainty from perturbative physics to be better than 5\%. This can be achieved by moving to a higher order in resummation accuracy (N3LL).  This is something we leave as a follow up to this paper.

In the SIDIS case we measure the transverse momentum imbalance between the groomed jet and the recoiling lepton. Once again we demand a jet mass measurement in order to ensure sensitivity to collinear physics only. A large part of the contribution to this comes from the soft and collinear radiation that lies outside the jet and, for low transverse momentum, probes the complete TMDPDF. The cross section is again presented to NNLL accuracy and involve much of the same ingredients as in the case of $e^+e^- \rightarrow$ dijets. A higher order perturbative calculation  is expected to reduce significatevely errors also in this case.

Concerning the hadronization effects we observe that grooming the jet allows us to have a wide angle jet, which is preferred in low energy experiments,  while still being free from non-global logarithms, which are non-factorizable  and they are usually present in un-groomed jets.  Nevertheless it is possible to measure directly the hadronization effects due to grooming. The idea  is to parametrize and extract all of the non-perturbative effects from $e^+e^- \rightarrow$ dijets and use them in SIDIS  since they contain all the same matrix elements (in addition to the TMDPDF) as explained in sec.~\ref{hadronization}. 
This gives us a robust way to access the TMDPDF while maintaining control over all other uninteresting  non-perturbative effects.

\section*{Acknowledgements}
The authors would like to thank Aditya Pathak, Iain W. Stewart, and Wouter J. Waalewijn for useful discussions. I.S. likes to acknowledge  the support of Los Alamos National Lab for his visit, during which part of the work was done. D.G.R. and   I.S. are supported by the Spanish MECD grant FPA2016-75654-C2-2-P. This project has received funding from the European Union Horizon 2020 research and innovation program under grant agreement No 824093 (STRONG-2020). D.G.R. acknowledges the support of the Universidad Complutense de Madrid through the predoctoral grant CT17/17-CT18/17. Y.M. and V.V. are supported by the U.S. Department of Energy through the Office of Science, Office of Nuclear Physics under Contract DE-AC52-06NA25396 and by an Early Career Research Award, through the LANL/LDRD Program. V.V. is also supported  within the framework of the TMD Topical Collaboration. L.Z. is supported by ERC grant ERC- STG-2015-677323.

\appendix
\section{Laplace and Fourier transformations}
\label{transforms}
We define the Fourier transform, $ \mathcal{FT}[f](\pmb{b})={f}(\pmb{b})$ of a function, $ f(\pmb{q}_T)= \mathcal{FT}^{-1}[{f}](\pmb{q}_T)$  as follows,
\begin{equation}
  {f}(\pmb{b}) = \int_{-\infty}^{+\infty} d \pmb{q}_T\; f(\pmb{q}_T) \exp(-i \pmb{b}\cdot \pmb{q}_T) \;,
\end{equation}
and the inverse transform
\begin{equation}
  f(\pmb{q}_T) = \int_{-\infty}^{+\infty}  \frac{d \pmb{b}\;}{(2\pi)^2}  {f}(\pmb{b})\exp(i \pmb{b}\cdot \pmb{q}_T) \;.
\end{equation}
In order to get the Fourier transforms of the plus distributions that appear in the factorization theorem we use,
\begin{equation}
  \label{eq:+expand}
  \frac{1}{(2\pi)\mu^2} \lp \frac{\mu^2}{q_T^2} \rp^{1+\alpha} = -\frac{1}{2\alpha} \delta^{(2)}(\pmb{q}_T) + \mathcal{L}_0(q_T^2,\mu^2) -\alpha  \mathcal{L}_1(q_T^2,\mu^2) + \mathcal{O}(\alpha^2) \;.
\end{equation}
Taking the Fourier transform of the left-hand-side (LHS) we get (see eq.~(E.2) of ref.~\cite{Chiu:2012ir})
\begin{align}
  \int_{-\infty}^{+\infty} \frac{d \pmb{q}_T}{(2\pi)}\;  \frac{1}{\mu^2} \lp \frac{\mu^2}{q_T^2} \rp^{1+\alpha}\exp(-i \pmb{b}\cdot \pmb{q}_T) 
  &
  = - \frac{e^{-2\alpha \gamma_E}}{2 \alpha}\frac{\Gamma(1-\alpha)}{\Gamma(1+\alpha)}\lp \frac{\mu} {\mu_E}  \rp^{2\alpha}
  \nn \\ &
  = - \frac{1}{2\alpha} + \ln\lp \frac{\mu} {\mu_E}  \rp +\alpha \ln^2\lp \frac{\mu} {\mu_E}  \rp +\mathcal{O}(\alpha^2)  \;,
\end{align}
where $\mu_E = 2 \exp(-\gamma_E)/b$ and $b\equiv \vert \pmb{b} \vert$ and in the second line we expanded in $\alpha$. Comparing this result with the RHS of eq.~(\ref{eq:+expand}) we get,
\begin{align}
  \label{eq:fourier}
  \mathcal{FT} \lb \delta^{(2)}(\pmb{q}_T)  \rb (\pmb{b}) &= 1 \;,\nn\\
  \mathcal{FT} \lb  \mathcal{L}_0(q_T^2,\mu^2) \rb (\pmb{b}) &=  \ln\lp \frac{\mu_E} {\mu}  \rp \;, \nn\\
  \mathcal{FT} \lb  \mathcal{L}_1(q_T^2,\mu^2) \rb (\pmb{b}) &=  \ln^2\lp \frac{\mu_E} {\mu} \rp \;. 
\end{align}
We define the convolution $f\otimes g$ with
\begin{equation}
  \lb f\otimes g \rb(\pmb{q}_T) = \int d \pmb{\ell}_{\perp} \;f(\pmb{q}_T - \pmb{\ell}_{\perp}) g(\pmb{\ell}_{\perp}) \;,
\end{equation}
such that
\begin{equation}
  \mathcal{F}\lb f\otimes g \rb (\pmb{b}) = {f}(\pmb{b}) \times {g}(\pmb{b}) \;. 
\end{equation}

Similarly for the distribution in the jet-thrust we often work in Laplace space where the corresponding convolutions translate to products. For these reason we define the Laplace transformation $\mathcal{LT}[f](s) = f(s)$ of jet-trust distribution $f(e) = \mathcal{LT}^{-1}[f](e) $ as follows:
\begin{equation}
  f(s) = \int_{-\infty}^{\infty} de\; \exp(-s e) f(e) \;,
\end{equation}
and the corresponding inverse transform
\begin{equation}
  {f}(e) = \frac{1}{2\pi i} \int_{\gamma-i \infty}^{\gamma+i \infty } ds\; \exp(s e) f(s) \;. 
\end{equation}
Similarly with the case of Fourier transform we use the following expansion to identify the Laplace transform of plus distributions that are present in the fixed order expansion of the jet and collinear-soft functions,
\begin{equation}
  \label{eq:+expand-e}
  \frac{1}{\xi} \lp \frac{\xi}{e}\rp^{1+\alpha} \Bvert_{e>0} = - \frac{1}{\alpha} \delta(e) + \mathcal{L}_0(e,\xi) -\alpha \mathcal{L}_1 (e,\xi) + \mathcal{O}(\alpha^2) \;,
\end{equation}
taking the Laplace transform of the LHS we get
\begin{equation}
  \label{eq:+expand-s}
  \int_0^{\infty} \frac{de}{\xi}\lp \frac{\xi}{e}\rp^{1+\alpha} \exp(-s e) = s^{\alpha} \Gamma(-\alpha) = -\frac{1}{\alpha} -  \ln ( \xi \tilde{s})  - \alpha \lp \frac{1}{2}  \ln^2 (\xi \tilde{s} )  + \frac{\pi^2}{12}\rp +\mathcal{O}(\alpha^2) \;,
\end{equation}
where $\tilde{s} \equiv s \exp(\gamma_E)$ and thus from comparing eq.~(\ref{eq:+expand-e}) and (\ref{eq:+expand-s}) we have
\begin{align}
  \mathcal{LT} \lb  \delta(e) \rb(s) &= 1  \;,\nn  \\
  \mathcal{LT} \lb  \mathcal{L}_0(e,\xi) \rb(s) &= - \ln ( \xi\tilde{s}) \;,\nn \\
  \mathcal{LT} \lb  \mathcal{L}_1(e,\xi) \rb(s) &=  \frac{1}{2}\ln^2 (\xi \tilde{s} ) + \frac{\pi^2}{12} \;.
\end{align}
\section{Operator definitions and one loop results}
\label{oneloop}

In this appendix we give the operator definitions of the factorization elements and their NLO expansions. From those we determine the renormalization functions, group equations, and corresponding anomalous dimensions. Many of the results presented here are already known and found in literature.

\subsection{Jet functions}
The quark and gluon jet function definitions, one loop calculation, and the corresponding Laplace transforms can be found in ref.~\cite{Frye:2016aiz}. Here we summarize their results. The quark jet function is given by,
\begin{equation}
  J_{q}(e,Q) = \frac{(2\pi)^3}{N_c} \text{tr} \; \Langle \frac{\slashed{\bar{n}}}{2} \chi_n (0) \delta (Q - \mathcal{P}^-) \delta^{(2)}(\pmb{\mathcal{P}}_{\perp}) \delta(e - \mathcal{E}) \bar{\chi}_n \Rangle \;,
\end{equation}
and the gluon
\begin{equation}
  J_{q}(e,Q) = \frac{(2\pi)^3}{N_c} \text{tr} \; \Langle \frac{\slashed{\bar{n}}}{2} \mathcal{B}^{\mu}_{n\perp} (0) \delta (Q - \mathcal{P}^-) \delta^{(2)}(\pmb{\mathcal{P}}_{\perp}) \delta(e - \mathcal{E}) \mathcal{B}_{n \perp \mu} \Rangle \;,
\end{equation}
where $N_c$ is the number of colors and $\mathcal{B}_{n \perp}^{\mu}$ is the gauge invariant gluon building block of the effective field theory,
\begin{equation}
  \mathcal{B}_{n \perp}^{\mu} = \frac{1}{g} [W_{n}^{\dag} (\mathcal{P}_{\perp}^{\mu} + g A_{n,\perp}^{\mu}) W_n] \;.
\end{equation}
As demonstrated earlier when working with the cumulant distribution (i.e., when integrating out to $\ecut$) it is useful to work in Laplace space.  The renormalized groomed jet function up to NLO contributions in Laplace space is given by
\begin{equation}
  \label{eq:jet}
  J_i(s,Q;\mu) = 1+ \frac{\alpha_s C_i}{2 \pi} \lbc L_J^2 +\bar{\gamma}_i L_J - \frac{\pi^2}{3} + c_i \rbc + \mathcal{O}(\alpha_s^2) \;,
\end{equation}
where for quark initiated jets we have
\begin{align}
  C_q &=C_F = \frac{N_c^2-1}{2N_c}\;,& \bar{\gamma}_q &=  \frac{3}{2}\;, & c_q &= \frac{7}{2}\;, 
\end{align}
and for gluon initiated jets we have
\begin{align}
  C_g &=C_A = N_c\;,& \bar{\gamma}_g &=  \frac{\beta_0}{2 C_A}\;, & c_g &= \frac{67}{18} - \frac{10}{9} \frac{n_f T_R}{C_A}\;, 
\end{align}
The logarithms, $L_J$ that appear in eq.~(\ref{eq:jet}) and the corresponding one loop anomalous dimensions are
\begin{align}
  L_J &= \ln \lp \frac{\mu^2 \tilde{s}}{Q^2}\rp \;, & \gamma^J =  \frac{\alpha_s C_i}{\pi} \lp 2 L_J + \bar{\gamma}_i \rp  + \mathcal{O}(\alpha_s^2) \;.
\end{align}
The anomalous dimension is defined through the RG equation satisfied by renormalized jet functions. In Laplace space this is 
\begin{equation}
  \frac{d}{d\ln \mu}  J_i(s,Q;\mu) =  \gamma^J (s,Q;\mu) J_i(s,Q;\mu) \;.
\end{equation}
In momentum space the above equation is written as convolution (in the invariant mass variable $e$), of the anomalous dimension and the renormalized jet function.

\subsection{Collinear-soft function}
\label{CSoft}
The operator definition of the invariant mass measurement collinear soft function is given by
\begin{equation}
  S_{cs}(e,Qz_{cut}) = \frac{1}{N_R} \text{tr} \Langle T\left(U_n^{\dagger}W_t\right)\mathcal{M}^{SD}_{e} \bar T \left(W_t^{\dagger} U_n\right) \Rangle \;,
\end{equation}
where $\mathcal{M}^{SD}_e$ is the invariant measurement function,
\begin{equation}
  \mathcal{M}^{SD}_e = \delta \left(e-(1- \Theta_{SD})\;\mathcal{E} \right) \;.
\end{equation}
Here we dropped the jet flavor (quark/anti-quark or gluon) for simplicity of notation and the normalization constant $N_R$ is simply the size of the representation for  SU($N_c$) of the $W_t$ and $U_n$ Wilson lines. For quark jets (fundamental  representation) we have $N_R =  N_c$ and for gluon jets (adjoint representation) we have $N_R = N_c^2 -1$. At NLO the bare collinear soft function is given by
\begin{equation}
  S_{cs,\text{bare}}(e,Q\zcut) =\delta(e)+ \frac{\alpha_s C_i}{\pi} \lbc - \frac{1}{\epsilon^2} \delta(e) + \frac{1}{\epsilon} \mathcal{L}_0(e,\xi) - \mathcal{L}_1(e,\xi) + \frac{\pi^2}{12} \delta(e)  \rbc + \mathcal{O}(\alpha_s^2) \;,
\end{equation}
where
\begin{equation}
  \xi \equiv \frac{\mu^2}{Q^2 \zcut}\;.
\end{equation}
Therefore we have for the renormalized function
\begin{equation}
  S_{cs}(e,Q\zcut) =\delta(e)+ \frac{\alpha_s C_i}{\pi} \lbc  - \mathcal{L}_1(e,\xi) + \frac{\pi^2}{12} \delta(e)  \rbc  + \mathcal{O}(\alpha_s^2) \;,
\end{equation}
where
\begin{equation}
  S_{cs,\text{bare}}(e,Q\zcut) = Z_{cs}\otimes S_{cs}(e,Q\zcut) \;,
\end{equation}
with
\begin{equation}
  Z_{cs}(e) =  \delta(e)+ \frac{\alpha_s C_i}{\pi} \lbc - \frac{1}{\epsilon^2} \delta(e) + \frac{1}{\epsilon} \mathcal{L}_0(e,\xi)  \rbc   + \mathcal{O}(\alpha_s^2) \;.
\end{equation}
In Laplace space for the renormalized collinear-soft function we get,
\begin{equation}
  S_{cs}(s,Q\zcut;\mu)  =1 - \frac{\alpha_s C_i}{2 \pi} L_{cs}^2   + \mathcal{O}(\alpha_s^2)\;,
\end{equation}
which  satisfies the following RGE
\begin{equation}
  \frac{d}{d\ln \mu} S_{cs}(s,Q\zcut;\mu) = \gamma^{cs}(s,\mu)  S_{cs}(s,Q\zcut;\mu) \;.
\end{equation}
The logarithm $L_{cs}$ and the corresponding anomalous dimension are 
\begin{align}
  L_{cs} &= \ln (\xi \tilde{s})\;,& \gamma^{cs}(s,\mu) &= - 2 \frac{\alpha_s C_i}{\pi}  L_{cs}   + \mathcal{O}(\alpha_s^2)\;.
\end{align}

\subsection{Soft function}
\label{GSoft}

The  soft function that appears in the factorization theorems in eq.~(\ref{eq:fact_ee}) and (\ref{eq:fact_ep}) is defined in eq.~(\ref{eq:soft}) and it has been calculated in several schemes at higher orders in QCD, as reported in sec.~\ref{sec:hierarchy}.
Here we report a one loop expression using the analytic regulator in momentum space,
\begin{multline}
  S_{\text{bare}}  = \delta^{(2)} (\pmb{q}_T) + \frac{ \alpha_s(\mu) C_i}{\pi} \lbc  \frac{4}{\eta} \lb \mathcal{L}_0(q_T^{2}, \mu^2 ) -\frac{1}{2 \epsilon} \delta^{(2)} (\pmb{q}_T)  \rb + \frac{1}{\epsilon} \lb \frac{1}{\epsilon} - 2 \ln \lp \frac{\nu}{\mu} \rp  \rb \delta^{(2)} (\pmb{q}_T) \\
  + 4 \mathcal{L}_0(q_T^{2},\mu^2) \ln \lp \frac{\nu}{\mu} \rp - 2 \mathcal{L}_1(q_T^{2},\mu^2) -\frac{\pi^{2}}{12} \delta^{(2)} (\pmb{q}_T)  
  \rbc  + \mathcal{O}(\alpha_s^2) .
\end{multline}
The renormalized soft function, $S$, is defined through
\begin{equation}
  S_{\text{bare}} = Z_{s}^{\perp}(\mu,\nu) \otimes  S (\mu,\nu) ,
\end{equation}
and satisfies the following renormalization group equations 
\begin{align}
  \frac{d}{d\ln{\mu}} S (\mu,\nu) &= \gamma^{s} (\mu,\nu)  S (\mu,\nu)\;,
  & \frac{d}{d\ln{\nu}} S (\mu,\nu) &= \gamma^{s}_{\nu} (\mu,\nu) \otimes S (\mu,\nu) \;.
\end{align}
Therefore we find for the one-loop corresponding impact parameter space quantities
\begin{equation}
  S (\mu,\nu) = 1 + \frac{ \alpha_s(\mu) C_i}{\pi} \lbc
  4 \ln\lp \frac{\mu_E} {\mu}  \rp \ln \lp \frac{\nu}{\mu} \rp - 2 \ln^2\lp \frac{\mu_E} {\mu}  \rp -\frac{\pi^{2}}{12}  
  \rbc  + \mathcal{O}(\alpha_s^2)  \;,
\end{equation}
\begin{equation}
  Z_{s}^{\perp}(\mu,\nu) = 1 + \frac{ \alpha_s(\mu) C_i}{\pi} \lbc  \frac{4}{\eta} \lb  \ln \lp \frac{\mu_E} {\mu}  \rp -\frac{1}{2 \epsilon}   \rb + \frac{1}{\epsilon} \lb \frac{1}{\epsilon} - 2 \ln \lp \frac{\nu}{\mu} \rp  \rb \rbc  + \mathcal{O}(\alpha_s^2) \;,
\end{equation}
with
\begin{align}
  \gamma^s (\mu,\nu) &= - 4 \frac{\alpha_s (\mu) C_i}{\pi} \ln \lp \frac{\nu} {\mu} \rp  + \mathcal{O}(\alpha_s^2)   \;, & 
  \gamma^{s}_{\nu} (\mu,\nu)&= 4 \frac{\alpha_s (\mu) C_i}{\pi} \ln \lp \frac{\mu_E} {\mu}  \rp   + \mathcal{O}(\alpha_s^2)\;.
\end{align}

The rapidity and renormalization scales used to produce our result are fixed using the $\zeta$-prescription \cite{Scimemi:2018xaf} adapted for this case. Later in the appendix we give a description of how one can use the rapidity regulated objects that have $\nu$ dependence to construct the subtracted rapidity divergences free objects but yet keep trace of the rapidity logs using the $\zeta$ parameter.


\subsection{Soft-collinear function}
\label{SoftC}

\begin{figure}[t!]
  \centerline{\includegraphics[width =  \textwidth]{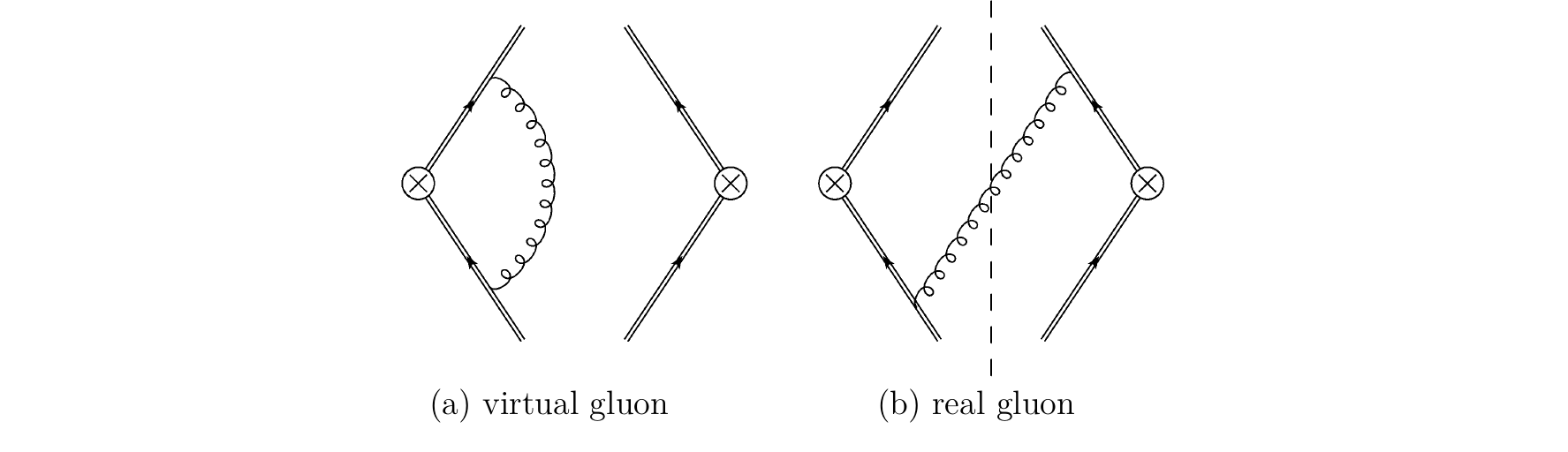}}
  \caption{The order $\mathcal{O}(\alpha_s)$ diagrams that contribute to the soft-collinear function.}
  \label{fig:soft_collinear}
\end{figure}

The soft-collinear function is defined by the matrix element 
\begin{equation}
  S_{sc}^{\perp}(Qz_{cut}) = \frac{1}{N_R} \text{tr} \Langle T\left(U_n^{\dagger}W_t\right)\mathcal{M}^{SD}_{\perp} \bar T \left(W_t^{\dagger} U_n\right) \Rangle \;, 
\end{equation}
and the groomed jet measurement function, $\mathcal{M}^{SD}_{\perp}$ is given in terms of the label momentum operator, $\mathcal{P}$,
\begin{equation}
  \label{eq:meas_sc}
  \mathcal{M}^{SD}_{\perp} = \Theta_{SD} \times \delta^2\left(\pmb{q}_T- \Theta_{SD}  \pmb{\mathcal{P}}_{\perp} \right) \;,
\end{equation}
where $\Theta_{SD}$ denotes the soft drop groomer. The collinear-soft modes only contribute to the invariant mass  measurement if they pass soft-drop, which is implemented by the $\Theta_{SD}$ term. The NLO calculation involves one real and one virtual diagram shown in fig.~\ref{fig:soft_collinear}. While the virtual diagram is scaleless.
The  diagram  with a real gluon needs to be integrated over the phase-space of soft gluon. This then yields non-vanishing contribution from when the soft gluon fails the grooming, 
\begin{equation}
  S_{sc,\text{NLO}}^{\perp} (Qz_{cut}) =  4 g^2 C_i \tilde \mu^{2\epsilon}\; \nu^{\eta}\; \int \frac{d^dk}{(2\pi)^{d-1}}\frac{\delta(k^2) \; \delta^{(2)}(\pmb{q}_T-\pmb{k}_{\perp})}{k^+ \; (k^-)^{1+\eta}}\theta( Q\zcut- k^{-}  ) \;.
\end{equation}
Performing the integrals we find for the bare quantity
\begin{multline}
  S_{sc,\text{bare}}^{\perp}(Q\zcut) = \delta^{(2)}(\pmb{q}_T)+\frac{\alpha_s C_i}{\pi} \lbc  - \frac{2}{\eta} \lb \mathcal{L}_0(q_T^2,\mu^2)-\frac{1}{2\epsilon} \delta^{(2)}(\pmb{q}_T)  \rb + \frac{1}{\epsilon} \ln\lp \frac{\nu}{Q\zcut} \rp \delta^{(2)}(\pmb{q}_T) \\- 2 \ln\lp \frac{\nu}{Q\zcut} \rp \mathcal{L}_0(q_T^2,\mu^2)\rbc   + \mathcal{O}(\alpha_s^2) \;,
\end{multline}
and for the renormalized quantity, $ S_{sc,}^{\perp}(Q\zcut;\mu,\nu)$ we have
\begin{equation}
  S_{sc,\text{bare}}^{\perp}(Q\zcut) =  Z_{sc}^{\perp}(\mu,\nu) \otimes S_{sc}^{\perp}(Q\zcut;\mu,\nu) \;,
\end{equation}
and satisfies the following renormalization group equations 
\begin{align}
  \frac{d}{d\ln{\nu}} S_{sc}^{\perp} (\mu,\nu) &= \gamma^{sc}_{\nu} (\mu,\nu) \otimes S_{sc}^{\perp} (\mu,\nu)
  \;, &
  \frac{d}{d\ln{\mu}} S_{sc}^{\perp} (\mu,\nu) &= \gamma^{sc} (\mu,\nu)  S_{sc}^{\perp} ( \mu,\nu) \;,
\end{align}
where the $Q\zcut$ dependence is suppressed to improve readability. In $\overline{\text{MS}}$ scheme the corresponding Fourier transform can be obtained using eq.~(\ref{eq:fourier}):
\begin{equation}
  \tilde{S}_{sc}^{\perp}(Q\zcut;\mu,\nu) =1+\frac{\alpha_s C_i}{\pi} \lbc - 2 \ln\lp \frac{\nu}{Q\zcut} \rp \ln\lp \frac{\mu_E} {\mu} \rp\rbc  + \mathcal{O}(\alpha_s^2) \;,
\end{equation}
\begin{equation}
  \tilde{Z}_{sc}^{\perp}(\mu,\nu) = 1 + \frac{\alpha_s C_i}{\pi} \lbc  - \frac{2}{\eta} \lb \ln\lp \frac{\mu_E} {\mu}\rp -\frac{1}{2 \epsilon}   \rb + \frac{1}{\epsilon} \ln\lp \frac{\nu}{Q\zcut} \rp  \rbc   + \mathcal{O}(\alpha_s^2) \;,
\end{equation}
and thus for the one-one-loop anomalous dimensions we get
\begin{align}
  \gamma_{\nu}^{sc}(\mu,\nu) &= -2 \frac{\alpha_s (\mu) C_i}{\pi} \ln \lp \frac{\mu_E}{\mu} \rp  + \mathcal{O}(\alpha_s^2) \;&  \gamma^{sc}(\mu,\nu) = 2 \frac{\alpha_s (\mu) C_F}{\pi}  \ln \lp \frac{\nu}{Q \zcut} \rp  + \mathcal{O}(\alpha_s^2) .
\end{align}


\section{Solution of renormalization group evolution equations}
\label{app:evolution}

In this appendix we discuss the solutions of both virtuality and rapidity renormalization group equations written in eq.~(\ref{eq:unmeasRG}). All factorization elements (hard, soft, soft-collinear, collinear-soft, and jet) satisfy renormalization group equations, but only transverse momentum dependent quantities have rapidity RGE.


\subsection{Renormalization group evolution}
\label{sec:RGEs}

The solution to the RGE in eq.~(\ref{eq:unmeasRG}) is
\begin{align}
  \label{eq:U}
  G(\mu)&= \mathcal{U}_G(\mu,\mu_0) G(\mu_0) \, , &\mathcal{U}_G(\mu,\mu_0)=\exp \left( K_G (\mu, \mu_0) \right) \left( \frac{\mu_0}{m_G} \right) ^{2 \;\omega_G(\mu, \mu_0)},
\end{align}
with
\begin{align}
  \label{eq:Kt}
  K_G(\mu, \mu_0) &= 2 \int_{\alpha (\mu_0)}^{\alpha(\mu)} \frac{d \alpha}{\beta[\alpha]} \Gamma^{G}[\alpha] \int_{\alpha(\mu_0)}^{\alpha}
  \frac{d \alpha'}{\beta[\alpha']} +\int_{\alpha (\mu_0)}^{\alpha(\mu)} \frac{d \alpha}{\beta[\alpha]} \Delta \gamma^G [\alpha] ,\\
  \label{eq:wt}
  \omega_G(\mu, \mu_0) &=  \int_{\alpha (\mu_0)}^{\alpha(\mu)} \frac{d \alpha}{\beta[\alpha]} \Gamma^{G} [\alpha].
\end{align}
Since in this work we are interested only in the NLL and NLL' result we may keep only the first two terms in the perturbative expansion of the cusp part (i.e., $\Gamma_0^G$, $\Gamma_{0}^{\text{cusp}}$, and $\Gamma_{1}^{\text{cusp}}$) and only the first term form the non-cusp part ($\gamma^{G}_{0}$). Performing this expansion we get,
\begin{align}
  \label{eq:K}
  K_G(\mu, \mu_0) &=-\frac{\gamma^G_0}{2 \beta_0} \ln r -\frac{2 \pi \Gamma^G_0}{(\beta_0)^2} \Big{\lbrack} \frac{r-1-r\ln r}{\alpha_s(\mu)}
  + \left( \frac{\Gamma_1^{\text{cusp}}}{\Gamma_0^{\text{cusp}}}-\frac{\beta_1}{\beta_0} \right) \frac{1-r+\ln r}{4 \pi}+\frac{\beta_1}{8 \pi \beta_0}
  \ln^2 r  \Big{\rbrack}, \\
  \label{eq:w}
  \omega_G(\mu, \mu_0) &= - \frac{\Gamma^G_0}{ 2 \beta_0} \Big{\lbrack} \ln r + \left( \frac{\Gamma^1_{\text{cusp}}}{\Gamma^0_{\text{cusp}}} -
  \frac{\beta_1}{\beta_0}  \right) \frac{\alpha_s (\mu_0)}{4 \pi}(r-1)\Big{\rbrack},
\end{align}
where $r=\alpha(\mu)/\alpha(\mu_0)$ and $\beta_n$ are the coefficients of the QCD $\beta$-function,
\begin{equation}
  \beta[\alpha_s] = \mu \frac{d \alpha_s}{d \mu}= -2 \alpha_s \sum_{n=0}^{\infty} \left( \frac{\alpha_s}{4 \pi} \right)^{1+n} \beta_n \; .
\end{equation}
The expressions for all ingredients necessary to perform the evolution of any function that appears in the factorization theorems we considered in this paper are given in tab.~\ref{tb:evolution}.
\begin{table}[t!]
  \renewcommand{\arraystretch}{1.4}
  \begin{center}
    \begin{tabular}{|c||c|c|c|c|c|}
      \hline
      Function          & $\Gamma^{G}_{0}$        & $\gamma^{G}_{0}$             & $m_{G}$                      \\\hline  \hline
      $H_{ij}$          &$-4(C_i+C_{\bar{j}})$     &$-4 \bar{\gamma}_i (C_i+C_j)$ & $Q$                                            \\\hline
      $S_{cs}$          &   $-8 C_i $             &   0                          & $Q\sqrt{\zcut/\tilde{s}}$                    \\\hline
      $J_i$             &   $ 8 C_i$             & $  4 \bar{\gamma}_i C_i$      &  $Q /\sqrt{\tilde{s}}$                \\\hline
      $B_{i/h}$         & 0                       &  $  4 \bar{\gamma}_i C_i + \gamma_{sc}^{0}$  & 0            \\\hline
      $S$      & $ 4(C_i+C_j)$           & 0                           & $\nu_s$                    \\\hline
      $S_{sc}^{\perp}$   & 0                      & $\gamma^{sc}_{0}$              &   n.a                     \\\hline
    \end{tabular}
    \caption{Anomalous dimensions coefficients for up to NLL accuracy: $\bar{\gamma}_q = 3/2$, $\bar{\gamma}_g = \beta_0/(2C_A)$, and $\gamma^{sc}_{0} = 2 \alpha_s (\mu) C_F / \pi  \ln (\nu/Q \zcut)$.}
    \label{tb:evolution}
  \end{center}
\end{table}
The coefficients for the expansion of the cusp anomalous dimension are 
\begin{align}
  \Gamma^{\text{cusp}}_{0} =& 4 C_F \;, \nn \\
  \Gamma^{\text{cusp}}_{1} =& 4 C_F \lb \lp \frac{67}{9}-\frac{\pi ^2}{3}\rp C_A -\frac{20 } {9} n_f T_R\rb \;, \nn \\
  \Gamma^{\text{cusp}}_{2} =& 4 C_F \lb \lp \frac{245}{6} - \frac{134}{27} \pi^2 + \frac{11}{45} \pi^4 + \frac{22}{3} \zeta_3 \rp C_A^2 + \lp -\frac{209}{108} + \frac{5}{27} \pi^2 - \frac{7}{3} \zeta_3  \rp 8 C_A n_f T_R  \nn \\
  &+ \lp 16 \zeta_3 -\frac{15}{3} C_F n_f T_R - \frac{64}{27} T_R^2 n_f^2 \rp  \rb \;.
\end{align}
The two loop non-cusp anomalous dimensions we need to NNLL RGEs are given by ref.~\cite{Frye:2016aiz}
\begin{align}
  \frac{1}{2}\gamma_{1}^{s} +   \gamma_{1}^{sc} &= \frac{C_i}{2} \lb 34.01 C_F + \lp \frac{1616}{27} -56 \zeta_3 - 9.31 \rp C_A - \lp \frac{448}{27} + 14.04 \rp n_f T_R - \frac{2}{3} \pi^2 \beta_0 \rb \;,\nn \\
  \gamma^{cs}_{1} =& C_i \lb -17.00 C_F + \lp -55.20 + \frac{22}{9} \pi^2 +56\zeta_3 \rp C_A + \lp 23.61 - \frac{8}{9}\pi^2 \rp n_f T_R \rb \nn \\
  \gamma^{q}_{1} =& C_F \lb  \lp 3 - 4 \pi^2 +48 \zeta_3 \rp C_F + \lp \frac{1769}{27}+\frac{22}{9} \pi^2 -80 \zeta_3  \rp C_A   \nn \\
  &+\lp -\frac{484}{27} - \frac{8}{9} \pi^2  \rp n_f T_R \rb \;, \nn \\
  \gamma_{1}^{s} +   \gamma_{1}^{sc} + \gamma_{1}^{B} = &  C_i \lb  \lp 20 -4 \pi^2 +48 \zeta_3 \rp C_F +  \lp 60.87 +\frac{22}{9}\pi^2 - 80 \zeta_3  \rp   C_A  \nn \\
  &+ \lp -24.94 -\frac{8}{9} \pi^2  \rp n_f T_R \rb \;.
\end{align}


\subsection{The connection between $\zeta$-parameter and rapidity regulator}
\label{sec:nu-zeta}
In the standard EFT approach one used the rapidity renormalization group (RRG) equations in order to resum large logarithms at the level of individual rapidity regulated terms \cite{Chiu:2011qc,Chiu:2012ir}. A more recent approach for performing the resummation of large logarithms in the TMD evolution it was introduced in ref.~\cite{Scimemi:2018xaf}. The approach is referred to as the $\zeta$-prescription. Here we rewrite the fixed order results using the rapidity regulator in the past sections in the form appropriate for implementing the $\zeta$-prescription. In the framework of ref.~\cite{Scimemi:2018xaf} one works with the rapidity divergent free quantity,
\begin{equation}
  S^{sub}_{sc}(b;\mu,\zeta) \equiv \sqrt{S^{\perp}_2(\bmat{b};\mu,\nu_s)} \; S_{sc}^{\perp}(\bmat{b},Q\zcut;\mu,\nu_{sc})\;,
\end{equation}
where we have explicitly show the dependence on the rapidity regulator parameters $\nu_s$ and $\nu_{sc}$. In the RRG approach this combination does not acquire rapidity evolution thus here in order to establish the rapidity evolution we fix the rapidity scales at two different values. Particularly we evaluate the soft-collinear rapidity scale at its canonical value, $\nu_{sc} = Q \zcut$, and we allow for the corresponding soft scale to float through a parameter $\zeta$: $\nu_s = \sqrt{\zeta}$.\footnote{Note that this is not a unique choice of scales since any choice for which $\nu_s/\nu_{sc} = \sqrt{\zeta} /(Q \zcut)$ will give the same result. } With this choice of scales we have,
\begin{equation}
  S^{sub}_{sc}(b;\mu,\zeta) = 1 + \frac{ \alpha_s(\mu) C_i}{2\pi} \lbc
  2 \ln\lp \frac{\mu_E} {\mu}  \rp \ln \lp \frac{\zeta}{\mu^2} \rp - 2 \ln^2\lp \frac{\mu_E} {\mu}  \rp -\frac{\pi^{2}}{12} 
  \rbc  + \mathcal{O}(\alpha_s^2)  \;.
\end{equation}
And according to the notation of eq.~(\ref{eq:dsevTMDs}) and (\ref{eq:dsevTMDs-z}) satisfies the following equations
\begin{align}
  \mu^2 \frac{d}{d\mu^2}  S^{sub}_{sc}(b;\mu,\zeta) &= \frac{1}{2} \gamma^{sc/sub}(\mu,\zeta)  S^{sub}_{sc}(b;\mu,\zeta)\;,&
  \zeta \frac{d}{d\zeta} S^{sub}_{sc}(b;\mu,\zeta) &= - \mathcal{D}(\mu)  S^{sub}_{sc}(b;\mu,\zeta) \;.
\end{align}
Its easy to show that the anomalous dimensions $\gamma^{sc/sub}$ and $\mathcal{D}$ are related to the RG and RRG anomalous dimensions of the global soft and soft-collinear function as follows,
\begin{equation}
  \gamma^{sc/sub}(\mu,\zeta) =  \frac{1}{2} \gamma^s + \gamma^{sc} = \Gamma^{\text{cusp}}[\alpha_s] \ln \lp\frac{\mu^2}{\zeta} \rp + \frac{1}{2} \Delta \gamma^s[\alpha_s] + \Delta \gamma^{sc}[\alpha_s] \;,
\end{equation}
and
\begin{equation}
  \mathcal{D}(\mu)  = \Gamma^{\text{cusp}}[\alpha_s] \ln \lp\frac{\mu}{\mu_E} \rp - \frac{1}{4} \Delta \gamma_\nu^s[\alpha_s] \;,
\end{equation}
where
\begin{align}
  \Delta\gamma_{\nu}^s = - \lp \frac{\alpha_s(\mu)}{4\pi}\rp^2  C_i \lb \lp  \frac{128}{9} - 56 \zeta_3 \rp C_A + \frac{112}{9} \beta_0  \rb + \mathcal{O}(\alpha_s^3).
\end{align}
It is easy to confirm by looking the above equations that the anomalous dimensions $\gamma^{sc/sub}$ and $\mathcal{D}$ satisfy the following differential equations,
\begin{align}
  \label{eq:dif-sub}
  \frac{d}{d\ln \zeta} \gamma^{sc/sub}(\mu,\zeta) &= - \Gamma_{\text{cusp}} \;, & \frac{d}{d\ln \mu} \mathcal{D}(\mu) &= + \Gamma_{\text{cusp}} \;.
\end{align}
Also comparing against the notation of eq.~(\ref{eq:unmeasRG}) we see that the non-cusp part, $\Delta\gamma^{sc/sub}$, of the anomalous dimension $\gamma^{sc/sub}$ is a linear combination of the corresponding non-cusp pieces of the global soft and soft-collinear functions. Particularly:
\begin{equation}
  \Delta\gamma^{sc/sub}(\mu) = \lp \frac{1}{2} \Delta \gamma^s[\alpha_s(\mu)] + \Delta \gamma^{sc}[\alpha_s(\mu)] \rp \;,
\end{equation}
and this statement is true to all orders in perturbative expansion.


\subsection{$\zeta$-prescription}

The implementation of the $\zeta$-prescription leads to the definition of optimal TMDs. We sketch here the procedure to obtain optimal TMDs referring to the original work~\cite{Scimemi:2018xaf} for further details. The anomalous dimensions $\gamma_F(\mu,\zeta)$ and $\mathcal{D}(\mu,b)$ governing the evolution can be thought as two components of a vector field in the plane ($\ln \mu^2,\,\ln \zeta$). The integrability condition, e.g. eq.~(\ref{eq:dif-sub}), states that such field is irrotational, i.e. locally conservative. This allows to define a scalar potential and guarantees that the evolution between two points in the ($\ln \mu^2,\,\ln \zeta$) space is independent of the path; in particular, no evolution occurs along equipotential lines. However, the perturbative expansion breaks the validity of such statement and in fact it was shown that numerical predictions largely depend on the choice of path. This limit is overcome by the \emph{improved} $\gamma$ \emph{solution}, that reinstates path-invariance by supplementing $\gamma_F$ with formally higher-order terms. If we let $F$ be a generic TMD, then the evolution kernel $R$, implicitly defined as
\begin{align} \label{eq:kernelDefinition}
  F(x,\pmb{b},\mu_f,\zeta_f) = R(\pmb{b}; \mu_f,\zeta_f;\mu_i,\zeta_i) F(x,\pmb{b},\mu_i,\zeta_i)\, ,
\end{align}
within the improved $\gamma$ solution yields
\begin{align} \label{eq:improvedGamma}
  R(\pmb{b}; \mu_f,\zeta_f;\mu_i,\zeta_i) = \exp\bigg\{
  \mathcal{D}(\mu_f,\pmb{b})\ln\Big(\frac{\mu_f^2}{\zeta_f}\Big)
  -\mathcal{D}(\mu_i,\pmb{b})\ln\Big(\frac{\mu_i^2}{\zeta_i}\Big)
  -\int_{\mu_i}^{\mu_f} \frac{d\mu}{\mu} 
  \big[2\mathcal{D}(\mu,\pmb{b}) + \gamma_V(\mu) \big]\bigg\}\, ,
\end{align}
where $\gamma_V$ is the noncusp anomalous dimension. 

Path independence allows one to apply the $\zeta$-\emph{prescription}, the key point of the method. The idea is setting the initial rapidity scale $\zeta_i = \zeta_{\mu_i}$ as a function of $\mu_i$ such that the scale-dependence of the initial TMDs vanishes independent of $\mu_i$. At one loop, this simply reads
\begin{align} \label{eq:oneLoopZetaPrescription}
  \zeta_\mu = e^{-\frac{\gamma_V}{\Gamma_{\mathrm{cusp}}}} \mu^2\, ,
\end{align}
and  the corrections to higher loops are evaluated in~\cite{Scimemi:2018xaf}.

\begin{figure*}[tb]
  \centering
  \includegraphics[width=0.7\textwidth]{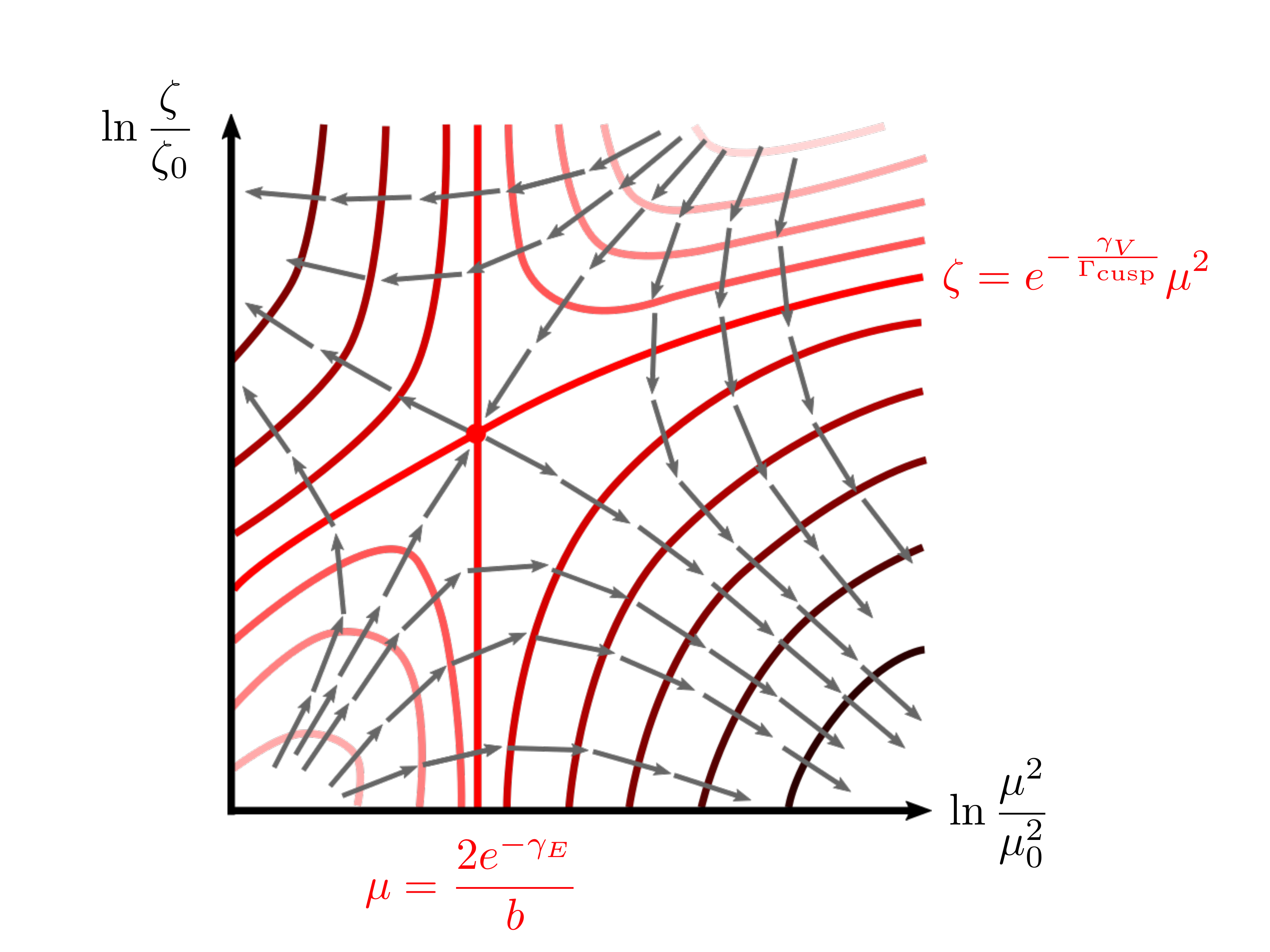}%
  \caption{Sketch of the geometry of the $(\zeta,\mu)$ plane where the double scale evolution takes place. The anomalous dimensions determine a conservative field (grey arrows) and the evolution is null among equipotential lines (shades of red). The intersection of two special equipotential lines (bright red) determines a saddle point; the zeta-prescription corresponds to running the evolution from this point, after reinstating path invariance. The equations for the special equipotential lines in the figure correspond to the one-loop result.}
  \label{fig:ZetaMuPlane}
\end{figure*}

The relation between $\zeta$ and $\mu$ draws a line in the $(\ln\zeta,\ln\mu^2)$ plane (\fig{ZetaMuPlane}). Since by requirement the TMDs are constant along it, this must be an equipotential line, which is well defined only if path-independence is restored. The remarkable fact with the $\zeta$ prescription is that, contrarily from standard evolution, the cancellation of large rapidity logarithms affecting the un-evolved TMDs is an internal mechanism. The rapidity evolution is still responsible for cancelling the large logarithms in the hard function, but the scale uncertainty of the evolution is now entirely decoupled from the definition of the TMDs (and in particular, from the non-perturbative model that enters their definition).

The definition of \emph{optimal TMDs} requires one more specification, which concerns the choice of initial scale $\mu_i$ (and consequently $\zeta_{\mu_i}$), and follows from TMD factorization. Considering TMD PDFs for definiteness, we have up to nonperturbative corrections
\begin{align}
  F_{a\leftarrow h} (x,\pmb{b},\mu,\zeta_\mu) = \sum_{b}\int_x^1 \frac{dy}{y}\, \mathcal{C}_{a\leftarrow b} (\tfrac{x}{y},\pmb{b},\mu,\zeta_\mu,\mu_{\mathrm{OPE}}) f_{b\leftarrow h} (y,\mu_{\mathrm{OPE}})\Big[ 1 + \mathcal{O}\big(b^2\Lambda_{\mathrm{QCD}}^2\big)\Big]\, ,
\end{align}
where $f_{b\leftarrow h}$ are the collinear PDFs, $\mathcal{C}_{a \leftarrow b}$ are transverse momentum matching coefficients known at two loop from ref.~\cite{Echevarria:2016scs}. The matching is performed at the scale $\mu_{\mathrm{OPE}}$. The choice of $\mu_{\mathrm{OPE}}$ is in general constrained by $\mu_i$, as they need to lie on the same half-plane with respect to the saddle point. This undesired feature is eliminated by choosing $\mu_i = \mu_{\mathrm{saddle}}$.

\subsection{Modeling of TMDPDF}
\label{sec:mod}

At small transverse momenta non-perturbative effects inside a TMD become dominant. A non-perturbative model valid for optimal TMDs was recently extracted in ref.~\cite{Bertone:2019nxa} by fitting combined data from Drell-Yan and Z-boson production. Since our groomed jet functions have the same rapidity evolution as the standard TMDs we will use the same model.
First, for large values of $b$ the initial scale, $\mu_{\mathrm{saddle}}$ enters the non-perturbative region. We correct for this by adopting the definition
\begin{align}
  \mu_i = \frac{2e^{-\gamma_E}}{b} + 2 \mbox{GeV}\, ,
\end{align}
which effectively imposes a higher cutoff on $b$. Second, the rapidity anomalous dimension is modified as follows,
\begin{equation}
  \mathcal{D}(\mu,\pmb{b}) = \mathcal{D}_{\mathrm{res}}(\mu,\pmb{b}^*) + c_0 b\, b^*\, ,
\end{equation}
where $c_0$ is a constant, the resummed anomalous dimension can be found at three loop in Refs.~\cite{Echevarria:2012pw,Scimemi:2018xaf}, and the b-star prescription is
\begin{align}
  \pmb{b}^* = \pmb{b}\Big(1 + \frac{b^2}{B_{\mathrm{NP}}^2}\Big)^{-\frac{1}{2}}\, .
\end{align}
The constants $c_0$ and $B_{\mathrm{NP}}$ specify the nonpertubative model in the case of $e^+e^- \to \dijets$. For SIDIS, additional input is required when building the TMD PDFs. Non-perturbative corrections to the factorization formula are modeled with a multiplicative, flavor-independent function $f_{\mathrm{NP}}$,
\begin{align}
  F_{a\leftarrow h} (x,\pmb{b},\mu,\zeta_\mu) = f_{\mathrm{NP}}(\pmb{b},x) \sum_{b} \int_x^1 \frac{dy}{y}\, \mathcal{C}_{a\leftarrow b} (\tfrac{x}{y},\pmb{b},\mu,\zeta_\mu,\mu_{\mathrm{OPE}}) f_{b\leftarrow h} (y,\mu_{\mathrm{OPE}})\, ,
\end{align}
whose explicit expression reads
\begin{align}
  f_{\mathrm{NP}} (\pmb{b},x) = \exp\bigg\{
  -\frac{\lambda_1 (1-x) + \lambda_2 x + \lambda_3 x (1-x)}{\sqrt{1+\lambda_4 x^{\lambda_5}b^2}}b^2\bigg\}\,,
\end{align}
generalizing the common choices of gaussian or exponential functions. The five parameters $\lambda_i$, together with $c_0$ and $B_{\mathrm{NP}}$, are listed in table 4 of ref.~\cite{Bertone:2019nxa} and were fitted within two different schemes: the first one treats them all as free parameters, while in the second one $B_{\mathrm{NP}}$ is fixed to 2.5~$\mbox{GeV}^{-1}$. The set of PDF used is NNPDF3.1~\cite{Ball:2017nwa}.

\section{IRC safety of the observable}
\label{app:IRCsafe}

\begin{figure}[t!]
  \centering
  \includegraphics[width= \textwidth]{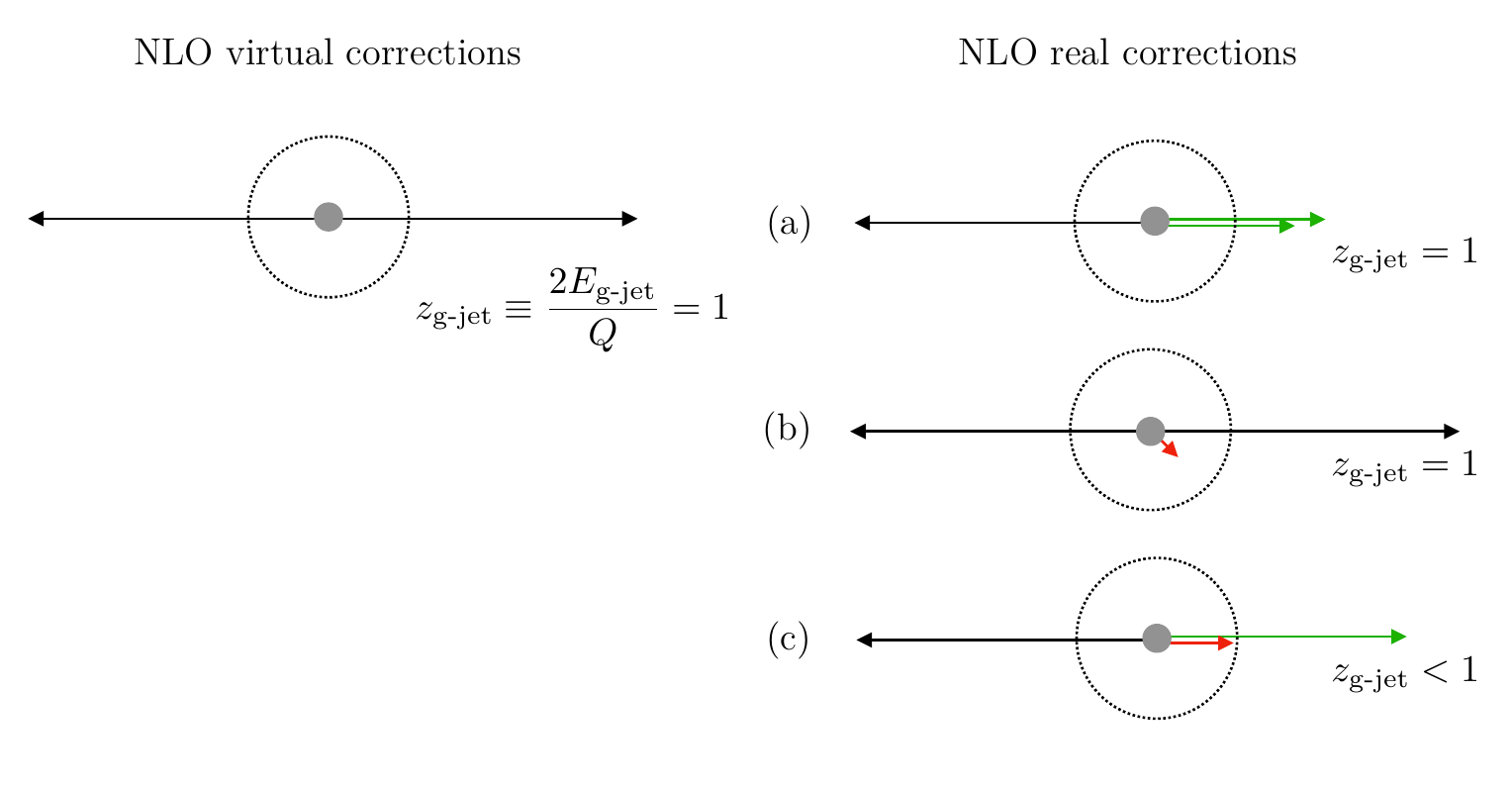}%
  \caption{IR divergent configurations for soft-drop groomed jets at $\mathcal{O}(\alpha_s)$.}
  \label{fig:IRC_safe}
\end{figure}

It is known that the energy difference between groomed and  ungroomed jet is an IRC unsafe quantity~\cite{Larkoski:2014wba, Larkoski:2015lea}. For the lepton-antilepton annihilation process it is trivial to show even at the leading non vanishing order, $\mathcal{O}(\alpha_s)$. In the standard jet cross section the collinear and soft divergence from the additional real gluon cancel against the  IR divergences  form virtual contribution. This is possible because in both the collinear and soft limits ($p_\text{gluon}^{0} \to 0$ and $p^{\mu}_{\text{gluon}} \parallel p^{\mu}_{q/\bar{q}} $ ) the energy of the (ungroomed) jet, $E_{\text{jet}}$, equals half of the center of mass energy, $Q$, i.e. $z_{\text{jet}}\equiv 2E_{\text{jet}} /Q =1$. In the case of groomed jets the phase space condition $p^{\mu}_{\text{gluon}} \parallel p^{\mu}_{q/\bar{q}}$ does not guarantee $z_{\text{g-jet}} = 1$ where $z_{\text{g-jet}}$ is the energy fraction fo the groomed jet, $z_{\text{g-jet}}\equiv 2E_{\text{g-jet}} /Q$. This is demonstrated in fig.~\ref{fig:IRC_safe} where configuration (a) corresponds to collinear gluon emission that passes soft-drop ($p^{\mu}_{\text{gluon}} \parallel p^{\mu}_{q/\bar{q}}$  and $p_{\text{gluon}}^{0} > \zcut Q$), (b) soft gluon emission ($p_{\text{gluon}}^{0} \to 0$), (c) collinear gluon emission that fails soft-drop ($p^{\mu}_{\text{gluon}} \parallel p^{\mu}_{q/\bar{q}}$  and $p_{\text{gluon}}^{0} < \zcut Q$). While divergences from the phase space  configurations (a) and (b)  can cancel against the virtual divergences, the ones in configuration (c) cannot.

This problem is usually solved when a jet substructure measurement (e.g. jet thrust $e$) is included. In this case the configuration (c) will only contribute to the $e=0$ bin and thus does not constitute a problem of IRC safety, if we constrain the result for finite values of the jet thrust.  In the observable we propose, the jet thrust measurement does not help since we require integrating over the range $e \in (0,\ecut)$, see for example eq.~(\ref{eq:int-ds}). However, the transverse momentum $\bmat{q}_T$ does since configuration (c) yields  only $\bmat{q}_T =0$ and we are interested only in finite values of the transverse momentum. Therefore in our proposed observable the $\bmat{q}_T$ measurement regulates the IRC divergences  in a similar manner to the (differential) jet-shape measurements.

Furthermore is important to notice that the quantity $\bmat{p}_{T\;\text{g-jet}} /z_{\text{g-jet}}$  does not directly relate to groomed jet energy, but rather to the groomed jet direction, which is collinear safe to all orders in $\alpha_s$,
\begin{equation}
  \frac{\bmat{p}_{T\;\text{g-jet}}}{z_{\text{g-jet}}} = Q\sin(\theta_{\text{g-jet}}) \bmat{n}_{T\;\text{g-jet}}
\end{equation}
where $\theta_{\text{g-jet}}$ is the angle between the groomed jet axis and the reference axis $\bmat{n}_{T\;\text{g-jet}}$ is the direction of the groomed jet with respect to the reference axis. Since the direction of the groomed jet is IR safe then the observable $\bmat{q}_{T}$ is as well.

\bibliographystyle{JHEP}
\normalbaselines 
\bibliography{bibliography} 

\end{document}